\documentclass[pdftex,twocolumn,epjc3]{svjour3}
\def\makeheadbox{\relax}

\RequirePackage{fix-cm}
\RequirePackage[colorlinks,citecolor=blue,urlcolor=blue,linkcolor=blue]{hyperref}
\smartqed  
\usepackage{graphicx}
\usepackage{mathptmx}

\usepackage{subcaption}
\usepackage[export]{adjustbox}
\usepackage{amsmath, amssymb, amsfonts}
\usepackage{xcolor}
\usepackage{verbatim}
\usepackage{mathtools}
\usepackage{siunitx}
\usepackage{enumitem}
\usepackage[
backend=biber,
style=numeric-comp,
sorting=none
]{biblatex}
\usepackage[]{hyperref}
\usepackage{booktabs}
\usepackage{tabularx}
\usepackage{siunitx}
\usepackage{bm}
\usepackage{lineno}

\graphicspath{{images/}{./}}

\numberwithin{equation}{section}
\numberwithin{figure}{section}
\numberwithin{table}{section}

\newcommand{\ra}[1]{\renewcommand{\arraystretch}{#1}}
\DeclareSIUnit\ele{\textit{e}}

\newcolumntype{L}[1]{>{\raggedright\let\newline\\\arraybackslash\hspace{0pt}}m{#1}}
\newcolumntype{C}[1]{>{\centering\let\newline\\\arraybackslash\hspace{0pt}}m{#1}}
\newcolumntype{R}[1]{>{\raggedleft\let\newline\\\arraybackslash\hspace{0pt}}m{#1}}

\AtBeginDocument{\RenewCommandCopy\qty\SI}

\begin{document}

\renewcommand{\hbar}{\mathchar'26\mkern-9mu h}


\title{Generating a highly uniform magnetic field inside the  magnetically shielded room of the n2EDM experiment
}

\author{
C. Abel\thanksref{Sussex}
\and
N.~J.~Ayres\thanksref{ETH}
\and
G.~Ban\thanksref{CAEN}
\and
G.~Bison\thanksref{PSI}
\and
K.~Bodek\thanksref{Cracow}
\and
V.~Bondar\thanksref{ETH}
\and
T.~Bouillaud\thanksref{LPSC, Bouillaud}
\and
D.C.~Bowles\thanksref{Kentucky}
\and
G.L.~Caratsch\thanksref{PSI,ETH}
\and
E.~Chanel\thanksref{Bern,ILL}
\and
W.~Chen\thanksref{PSI,ETH}
\and
P.-J.~Chiu\thanksref{PSI,ETH,Chiu}
\and
C.~Crawford\thanksref{Kentucky}
\and
B.~Dechenaux\thanksref{CAEN}
\and
C.B.~Doorenbos\thanksref{PSI,ETH}
\and
S.~Emmenegger\thanksref{ETH}
\and
L.~Ferraris-Bouchez\thanksref{LPSC}
\and
M.~Fertl\thanksref{Mainz}
\and
P.~Flaux\thanksref{CAEN}
\and
A.~Fratangelo\thanksref{Bern}
\and
D.~Goupillière\thanksref{CAEN}
\and
W. C. Griffith\thanksref{Sussex}
\and
Z. Grujic\thanksref{Serbia}
\and
D.~H\"ohl\thanksref{PSI,ETH}
\and
M.~Kasprzak\thanksref{Leuven,Kasprzak}
\and
K.~Kirch\thanksref{PSI,ETH}
\and
V.~Kletzl\thanksref{PSI,ETH}
\and
S.V.~Komposch\thanksref{PSI,ETH}
\and
P.~A.~Koss\thanksref{Leuven,Koss}
\and
J.~Krempel\thanksref{ETH}
\and
B.~Lauss\thanksref{PSI}
\and
T.~Lefort\thanksref{CAEN, Lefort}
\and
A.~Lejuez\thanksref{CAEN}
\and
R.~Li\thanksref{Leuven}
\and
M.~Meier\thanksref{PSI}
\and
J.~Menu\thanksref{LPSC}
\and
K.~Michielsen\thanksref{LPSC}
\and
P.~Mullan\thanksref{ETH}
\and
A.~Mullins\thanksref{Kentucky}
\and
O.~Naviliat-Cuncic\thanksref{CAEN}
\and
D.~Pais\thanksref{PSI,ETH}
\and
F.M.~Piegsa\thanksref{Bern}
\and
G.~Pignol\thanksref{LPSC, Pignol}
\and
G.~Quemener\thanksref{CAEN}
\and
M.~Rawlik\thanksref{ETH}
\and
D.~Rebreyend\thanksref{LPSC}
\and
I.~Rienaecker\thanksref{PSI}
\and
D.~Ries\thanksref{PSI}
\and
S.~Roccia\thanksref{LPSC}
\and
D.~Rozpedzik\thanksref{Cracow}
\and
A.~Schnabel\thanksref{PTB}
\and
P.~Schmidt-Wellenburg\thanksref{PSI}
\and
E.P.~Segarra\thanksref{PSI}
\and
N.~Severijns\thanksref{Leuven}
\and
C.A.~Smith\thanksref{Kentucky}
\and
K.~Svirina\thanksref{LPSC,ILL}
\and
R.~Tavakoli\thanksref{Leuven,Tavakoli}
\and
J.~Thorne\thanksref{Bern}
\and
S.~Touati\thanksref{LPSC}
\and
J.~Vankeirsbilck\thanksref{Leuven}
\and
R.~Virot\thanksref{LPSC}
\and
J.~Voigt\thanksref{PTB}
\and
E.~Wursten\thanksref{Leuven,Wursten}
\and
N.~Yazdandoost\thanksref{Mainz2}
\and
J.~Zejma\thanksref{Cracow}
\and
N.~Ziehl\thanksref{ETH}  
\and
G.~Zsigmond\thanksref{PSI}
}

\thankstext{Bouillaud}{Corresponding author: thomas.bouillaud@epp.phys.kyushu-u.ac.jp.}
\thankstext{Lefort}{Corresponding author: lefort@lpccaen.in2p3.fr.}
\thankstext{Pignol}{Corresponding author: guillaume.pignol@lpsc.in2p3.fr}
\thankstext{Kasprzak}{Present address: Paul Scherrer Institute, Villigen, Switzerland}
\thankstext{Koss}{Present address: Fraunhofer Institute for Physical Measurement Techniques, 79110 Freiburg, Germany}
\thankstext{Tavakoli}{Present address: IMO-IMEC, Campus Diepenbeek, Agoralaan Gebouw D, 3590 Diepenbeek, Belgium}
\thankstext{Wursten}{Present address: CERN, Esplanade des Particules 1, 1217 Meyrin, Switzerland, and RIKEN, Fundamental Symmetries Laboratory, 2-1 Hirosawa, Wako, Saitama 351-0198, Japan}
\thankstext{Chiu}{Present address: University of Zurich, Zurich, Switzerland}
\thankstext{ILL}{Present address: Institut Laue Langevin, Grenoble, France}
\institute{
    Department of Physics and Astronomy,      University of Sussex, Falmer, Brighton BN1 9QH, UK \label{Sussex}
    \and
    ETH Zürich, Institute for Particle Physics and Astrophysics, CH-8093 Zürich, Switzerland \label{ETH}
    \and
     Université de Caen Normandie, ENSICAEN, CNRS/IN2P3, LPC Caen UMR6534, F-14000 Caen, France\label{CAEN}
    \and
    Institute for Nuclear and Radiation Physics, KU Leuven, B-3001 Leuven, Belgium\label{Leuven}
    \and
    Laboratory for Particle Physics, PSI Center for Neutron and Muon Sciences, Paul Scherrer Institut,  Forschungsstrasse 111, 5232 Villigen PSI, Switzerland\label{PSI}
    \and
    Marian Smoluchowski Institute of Physics, Jagiellonian University, 30-348 Cracow, Poland \label{Cracow}
    \and
    LPSC, Université Grenoble Alpes, CNRS/IN2P3, Grenoble, France\label{LPSC}
    \and
    University of Bern, Albert Einstein Center for Fundamental Physics, CH-3012 Bern, Switzerland\label{Bern}
    \and
    University of Kentucky, Lexington, USA\label{Kentucky}
    \and
    Institute of Physics, Johannes Gutenberg University, D-55128 Mainz, Germany\label{Mainz}
    \and
    Institute of Physics Belgrade, University of Belgrade, 11080 Belgrade, Serbia \label{Serbia}
    \and
    Department of Chemistry - TRIGA site, Johannes Gutenberg University, 55128 Mainz, Germany\label{Mainz2}
    \and
    Physikalisch Technische Bundesanstalt, Berlin, Germany\label{PTB}
}

\date{June 2024}

\maketitle

\begin{abstract}
We present a coil system designed to generate a highly uniform magnetic field for the n2EDM experiment at the Paul Scherrer Institute. 
It consists of a main $B_0$ coil and a set of auxiliary coils mounted on a cubic structure with a side length of $\qty{273}{cm}$, inside a large magnetically shielded room (MSR). 
We have assembled this system and characterized its performances with a mapping robot. 
The apparatus is able to generate a $\qty{1}{\micro T}$ vertical field with a relative root mean square deviation $\sigma(B_z)/B_z = \num{3e-5}$ over the volume of interest, a cylinder of radius $\qty{40}{cm}$ and height $\qty{30}{cm}$. 
This level of uniformity overcomes the n2EDM requirements, allowing a measurement of the neutron Electric Dipole Moment with a sensitivity better than $\qty{1e-27}{\ele.cm}$. 


\end{abstract}

\section{Introduction} \label{sec_intro}

n2EDM is an apparatus connected to the ultracold neutron source at the Paul Scherrer Institute\cite{lauss2021ucn, lauss2021bis}, designed to measure  the electric dipole moment (EDM) of the neutron $d_n$ with a sensitivity better than $\qty{1e-27}{\ele.cm}$ \cite{n2edm}.
This represents an order of magnitude improvement compared to the previous version of the experiment, which set the best upper limit to date on $d_n$ \cite{nedm}.
For a discussion on the landscape of current and future experiments searching for non\hyphen  zero EDMs of subatomic particles, and their role as sensitive probes of new physics beyond the Standard Model, we refer to the recent articles \cite{Chupp2019, Cairncross2019, alarcon2022electric}.




\begin{figure}[ht!]
    \centering
    \includegraphics[width=0.49\textwidth]{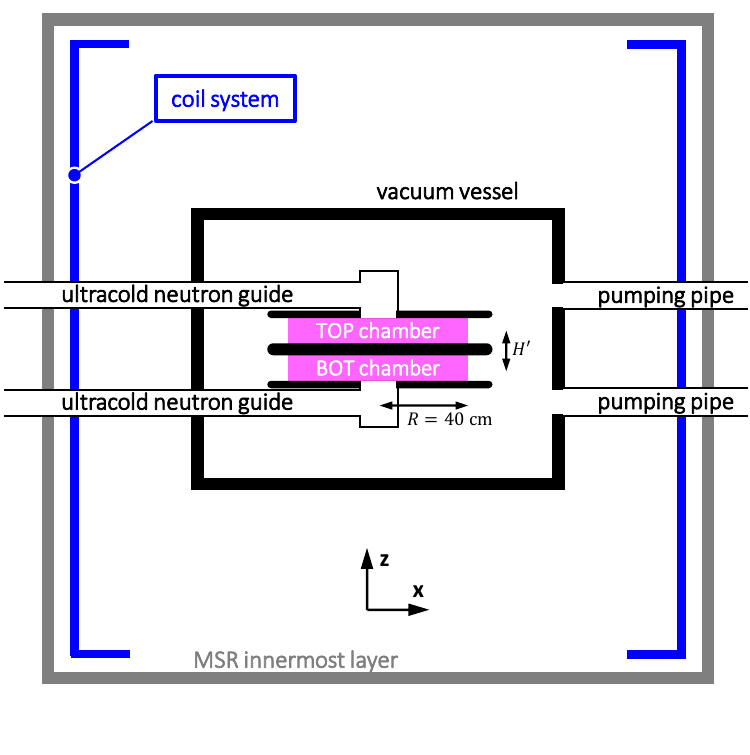}
    \includegraphics[width=0.49\textwidth]{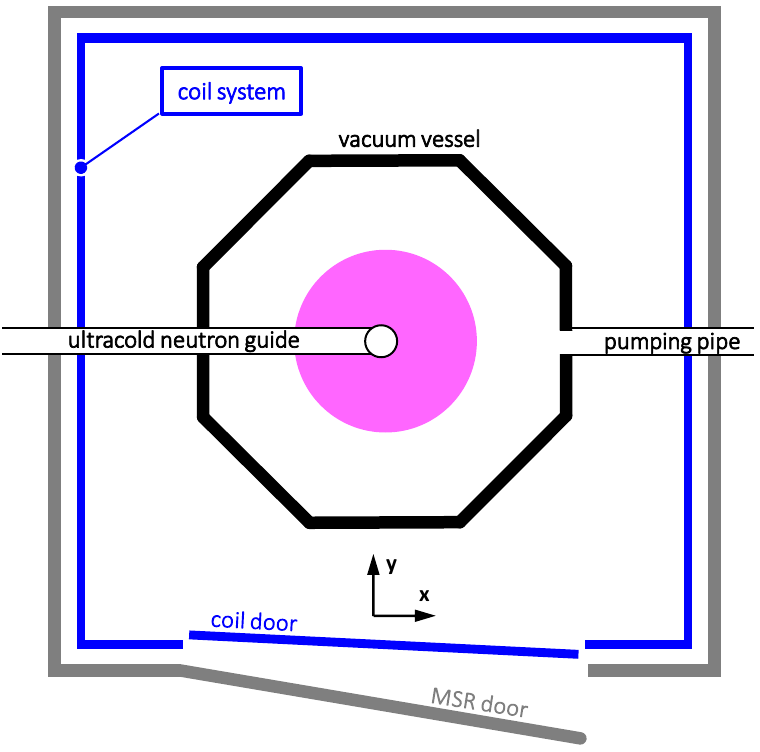}
    \caption{Schematic depiction of the n2EDM apparatus inside the Magnetic Shieding Room (MSR), view from a vertical cut (top figure) and a horizontal cut (bottom figure). The coordinate system is defined such that the $y$ axis points from the MSR door to the back of the MSR in the horizontal plane. The MSR together with the coil system (in blue) are designed to generate a uniform vertical field inside the MSR volume, and especially so inside the double precession chamber volume (in pink).}
    \label{fig_n2EDM-drawing}
\end{figure}

Figure \ref{fig_n2EDM-drawing} shows a scheme of n2EDM relevant for the present article.
In the experiment, spin-polarized ultracold neutrons and ${}^{199}\text{Hg}$ atoms will be stored for several minutes in two large precession chambers.
Each chamber has a cylindrical shape of radius $R=40$~cm and height $H=12$~cm.
The chambers are stacked vertically, with a height separation of $H'=18$~cm between their respective centers.
During storage, the neutrons and mercury atoms will be exposed to
(i) a strong vertical electric field, $E=\qty{15}{kV/cm}$, of opposite polarity in the two chambers, and
(ii) a weak static vertical magnetic field, ideally identical in the two chambers.
In a first phase of the experiment, the magnetic field will be set to the \emph{baseline value} of \SI{1}{\micro T}, as in the previous single-chamber experiment \cite{nedm}.
In a second phase, the field will be set to the so-called \emph{magic value} of $\qty{10}{\micro T}$ intended to suppress the main systematic effect \cite{magic}.


The particle spins will precess about the fields due to their magnetic (and possibly non\hyphen  zero electric) dipole moments.
The neutron precession frequency will be measured using Ramsey's technique of separated rotating fields \cite{Ramsey}, while the mercury precession frequency will be read-out optically during the precession.
The (possibly non\hyphen  zero) EDM of the neutron will cause a tiny difference in the neutron precession frequency upon reversal of the electric field.
The mercury atoms are used as a co-magnetometer: the atoms average the magnetic field in essentially the same volume and during the same time as the neutrons.
In addition, an array of 112 cesium atomic magnetometers placed around the chambers will be used for the online control of the uniformity of the magnetic field.

A stable and uniform magnetic field has to be generated in a large volume encompassing the stacked precession chambers.
The large volume of the chambers ($\times 6$ compared to the previous experiment) allows an increase in the number of stored neutrons and therefore a boost of the statistical sensitivity \cite{n2edm}.
The stack is placed in a nonmagnetic vacuum vessel, which is itself installed in a Magnetically Shielded Room (MSR).
The MSR \cite{MSR} is a cubic structure of six ferromagnetic layers, with an interior volume of side length 293~cm.
In addition, the passive MSR is complemented by an Active Magnetic Shield \cite{AMS} with feedback-controlled coils external to the MSR (not shown in figure \ref{fig_n2EDM-drawing} as it is not directly relevant to the present subject).

In this article, we present the coil system designed and built to generate the magnetic field inside the MSR.
In section \ref{sec_uniformity}, we discuss the requirements for the field generation, in particular about the uniformity.
Then, in section \ref{sec_design}, we lay out the detailed design of the coil system.
Finally, in section \ref{sec_characterization}, we report on the results of a magnetic field mapping campaign characterizing the performances of the system at the baseline value for the magnetic field, $B_0 = \qty{1}{\micro T}$.

\section{Magnetic field uniformity in n2EDM} \label{sec_uniformity}
\label{Requirements}

The requirements related to magnetic field uniformity are expressed in a convenient field parametrization, of the form
\begin{align}
    \mathbf{B}(\mathbf{r}) 
    &= \sum_{l\geq 0}^{}\sum_{m=-l-1}^{l+1} G_{lm} \bm{\Pi}_{lm}(\mathbf{r}) \label{eq_harmonic-expansion-old} \\
    &= \sum_{l\geq 0}^{}\sum_{m=-l-1}^{l+1} \frac{\Acute{G}_{lm}}{D_{l}^{l-1}}\bm{\Pi}_{lm}(\mathbf{r}). \label{eq_harmonic-expansion}
\end{align}
The first parametrization (equation \eqref{eq_harmonic-expansion-old}) was introduced in \cite{uniformity} and is referred to as the \textit{harmonic expansion}. In the above equations, the harmonic modes $\bm{\Pi}_{lm}(\mathbf{r})$ are polynomial functions of degree $l$ which are determined explicitly by requiring that the field satisfies Maxwell's stationary equations $\boldsymbol{\nabla}\cdot \mathbf{B} = 0$ and $\boldsymbol{\nabla}\times \mathbf{B} = 0$. A table of those polynomials along with their visual representations in the transverse plane can be found in \ref{sec_harmonic}. The coefficients of the expansion $G_{lm}$ are generalized magnetic gradients, usually expressed in units of $\unit{pT/cm^l}$. However, it is more convenient to compare normalized magnetic gradients $\Acute{G}_{lm}$, with units of $\unit{pT/cm}$, that we introduce in a new parametrization (equation \eqref{eq_harmonic-expansion}). To this end we define normalizing distances $D_{l}$, in units of $\unit{cm}$, which are determined by the geometry of the precession chambers through the normalization detailed in \ref{sec_norm}. Their numerical values are specified in table \ref{table_D}. In n2EDM the expansion is carried out up to order $l=7$ because, as we will later show, the systematic effect generated by terms of order $l=7$ and beyond is negligible.

\begin{table}[ht!]
\centering
\begin{tabular}{@{\kern\tabcolsep} l ccccccc @{\kern\tabcolsep}}
    \toprule
    $l$ & $1$ & $2$ & $3$ & $4$ & $5$ & $6$ & $7$\\
    \midrule
    $D_{l}$ $(\text{cm})$ &  $1$ & $18$ & $23.7$ & $-29.1$ & $31.8$ & $39.7$ & $33.8$\\
    \bottomrule
\end{tabular}
\caption{Normalizing distances of the harmonic expansion, up to $l=7$.}
\label{table_D}
\end{table}

\subsection{Uniformity requirements related to statistical sensitivity}
\label{Stat_req}

Magnetic field uniformity has a strong influence on the statistical sensitivity of the neutron precession frequency measurement in n2EDM, and is constrained by two requirements \cite{n2edm}.

The first of these concerns the decay of the neutrons' spins polarization during a Ramsey cycle, which should be kept minimal in order to maximize the statistical sensitivity. Non\hyphen   uniformities in the vertical magnetic field component
lead to a depolarization of the neutrons' spins. One can show that the decay rate of the transverse polarization
\begin{equation}
    \frac{1}{T_2} =  \gamma_n^2 \sigma^2(B_z)\tau_c, \label{eq_depolarization}
\end{equation}
described by spin-relaxation theory \cite{Redfield}, is proportional to the root mean square of the spatial field variations $\sigma(B_z) = \sqrt{\left<(B_z-\left<B_z\right>)^2\right>}$, where the angle brackets indicate an average over the precession volume. In equation \eqref{eq_depolarization}, $\gamma_n$ is the neutron's gyromagnetic ratio and $\tau_c$ the autocorrelation time of UCN motion. The latter was originally determined in the nEDM experiment by measuring the transverse depolarization in the presence of a large applied gradient \cite{uniformity}. Based on this measurement, we extrapolated the value of $\tau_c$ to account for the increased diameter of the chambers used in the current experiment, to $\tau_c=\qty{120}{ms}$ \cite{n2edm}. In the design, we impose that the neutron spin polarization must not decrease by more than $2\%$ after $\qty{180}{s}$ of precession.
This translates to a requirement on the vertical non\hyphen  uniformity inside each precession chamber
\begin{equation}
    \sigma(B_z) < \qty{170}{pT}. \label{eq_uniformity}
\end{equation}

The second requirement is due to the double chamber configuration of n2EDM, where the presence of a magnetic gradient between the two chambers does not allow one to simultaneously measure the top and bottom precession frequencies at optimal sensitivity. Specifically, since the rotating field responsible for the Ramsey spin flip is applied to both chambers simultaneously, its frequency, $f_{\text{RF}}$, should be set to a value that minimizes the statistical error of the precession frequency extraction from the Ramsey curves of both chambers.
As the two resonance curves shift with the vertical magnetic field, we require that the vertical gradient between the two chambers, defined as the \textit{top-bottom gradient} $G_{\text{TB}}=\left(\left<B_z\right>_\text{TOP}-\left<B_z\right>_\text{BOT}\right)/H'$, remains below a value that corresponds to an imposed $2\%$ loss in sensitivity. This condition is known as the \textit{top-bottom resonance matching condition}
\begin{equation}
    |G_{\text{TB}}| < \qty{0.6}{pT/cm}. \label{eq_gtb}
\end{equation}

The coil system of n2EDM is designed to generate a uniform magnetic field that satisfies both of these requirements.

\subsection{The false neutron EDM due to non\hyphen uniformities}

The largest systematic error in n2EDM is a shift in the neutron to mercury spin precession frequency ratio due to a spin-relaxational effect experienced by mercury atoms. This effect is detailed extensively in section $4$ of the n2EDM design article \cite{n2edm}. Magnetic non\hyphen  uniformities $B_z-\left<B_z\right>$ in combination with a relativistic motional field $v\times E/c^2$ shift the precession frequencies of both neutrons and mercury atoms, but more so of mercury atoms. Since the neutron EDM $d_n$ is extracted from the ratio $\mathcal{R}=f_n/f_\text{Hg}$ of the two measured frequencies, the shift in the precession frequencies of the mercury atoms generates an error on the EDM measurement, referred to as the \textit{false EDM} and denoted $d_{n\leftarrow\text{Hg}}^{\text{false}}$. 
As we seek to measure $d_n$ at a sensitivity of $\qty{e-27}{\ele.cm}$, we require that the false EDM is kept below:
\begin{equation}
    d_{\text{n}\leftarrow\text{Hg}}^{\text{false}} < \qty{3e-28}{\ele.cm}. \label{eq_falseEDM-condition}
\end{equation}

The issue of precession frequency shifts that generate false EDM signals have been extensively studied in the past decades \cite{1996-Lamoreaux, 2004-Pendlebury, 2005-Lamoreaux-Golub, 2006-Lamoreaux, 2011-Clayton, 2012-Swank-Golub, pignol2012, Pignol_2015, 2015-Golub, 2016-Golub, spinrelaxPignol}. 
In the case of mercury atoms with a thermal ballistic motion inside a low magnetic field (valid for the n2EDM \SI{1}{\micro T} field), the false EDM is written \cite{n2edm}
\begin{align}
    d_{n\leftarrow \text{Hg}}^{\text{false}} &= -\frac{\hbar\left|\gamma_n \gamma_{\text{Hg}}\right|}{2c^2} \left<\rho B_\rho\right> \label{eq_falseEDM-geometry} \\
    &= \frac{\hbar\left|\gamma_n \gamma_{\text{Hg}}\right|}{2 c^2}\frac{R^2}{4}\left(G_{\text{TB}} + \Acute{G}_{3 \, 0} + \Acute{G}_{5 \, 0} + \hdots\right) \label{eq_falseEDM-gradients} \\
    &= \frac{\qty{1.26e-26}{\ele.cm}}{\unit{pT/cm}}\times \left(G_{\text{TB}} + \Acute{G}_{3 \, 0} + \Acute{G}_{5 \, 0} + \hdots\right), \nonumber
\end{align}
where $\rho$ is the radial coordinate in the transverse plane and $B_\rho$ the radial field component. The angle brackets in the first equality indicate a volume average over the two precession chambers.
The second equality is obtained by deploying the harmonic magnetic field expansion \eqref{eq_harmonic-expansion} and is specific to the geometry of n2EDM. Equation \eqref{eq_falseEDM-gradients} tells us that the false EDM is proportional to a particular set of magnetic gradients. This is due to the double chamber geometry, for which only $l$-odd, $m=0$ harmonic modes yield a non\hyphen zero false EDM. We divide these modes into two categories: (i) those visible in the online monitoring of n2EDM, in that they generate a top-bottom gradient $G_{\text{TB}}$, and 
(ii) those that generate a false EDM while satisfying $G_{\text{TB}}=0$, and are therefore not fully accounted for by the online analysis. The latter are for this reason referred to as \textit{phantom modes}.

Because of equation \eqref{eq_falseEDM-gradients}, the systematical requirement \eqref{eq_falseEDM-condition} is effectively a requirement on the uniformity of the n2EDM magnetic field. There are two strategies to control this systematic effect in n2EDM. 

\sloppy The first is to simply measure the magnetic non\hyphen  uniformities involved in equation \eqref{eq_falseEDM-gradients} and estimate the false EDM. These measurements should be accurate enough so that $\delta d_{\text{n}\leftarrow\text{Hg}}^{\text{false}} < \qty{3e-28}{\ele.cm}$. While the top-bottom gradient $G_{\text{TB}}$ is accurately monitored online, the phantom gradients $\Acute{G}_{2k+1\,0}$ are measured offline with the n2EDM mapper. This imposes an additional requirement on the reproducibility of the phantom modes. 

The second strategy is to make the n2EDM magnetic field more uniform in order to suppress the false EDM below its systematic requirement. This field optimization strategy is implemented thanks to a set of auxiliary coils designed to target specific harmonic modes, particularly the $l$-odd, $m=0$ modes responsible for the false EDM.

The success of both strategies relies on the fulfillment of two common conditions. The core systematical requirement \eqref{eq_falseEDM-condition} translates to a requirement on (i) the reproducibility of the generated phantom modes before and during data-taking \\ $\sigma(\Acute{G}_{3\,0}), \sigma(\Acute{G}_{5\,0}), \sigma(\Acute{G}_{7\,0}) < \qty{23}{fT/cm}$, and (ii) the accuracy of the offline measurement of these modes $\delta\Acute{G}_{3\,0}, \delta\Acute{G}_{5\,0}, \delta\Acute{G}_{7\,0} < \qty{23}{fT/cm}$. We also note that while there still is a way of monitoring the third (and possibly fifth) degree phantom mode(s) online thanks to a cesium magnetometer array, as detailed in \cite{n2edm}, a redundant measurement of the phantom modes with the mapper is crucial to the control of such a debilitating systematic effect.
A summary of both statistical and systematical requirements and corresponding measurements is given in table \ref{table_requirements}.

\section{Design of the n2EDM coil system} \label{sec_design}

The inner coil system of the experiment consists of the main $B_0$ coil, an array of 56 independent optimization coils, and seven specific coils referred to as ``gradient coils" \cite{n2edm}. The $B_0$ field is generated by the single $B_0$ coil and the induced magnetization of the MSR innermost layer. The 56 independent optimization coils are used to cancel the remaining field non\hyphen  uniformities. Finally, the gradient coils generate specific magnetic gradients that play an important role in the measurement procedure. The coil system was designed so that it can produce a $B_0$ field of \SI{1}{\micro T}, as in the previous experiment \cite{nedm}, or \SI{10}{\micro T}, the magic field value which will be used in a second phase \cite{magic}. 



\subsection{The $B_0$ coil design}


The main part of the $B_0$ coil is a vertical square solenoid installed within the innermost chamber of the MSR. Infinite solenoids generate uniform magnetic fields. This statement remains true for finite solenoids inserted in a high permeability material. The coupling between the solenoid and the shield mimics an infinite solenoid on the condition that the coil extremities are closed off by two high permeability material planes perpendicular to the solenoid axis. Both planes define perfect boundary conditions for the vertical field component. The solenoid was therefore designed as long and as wide as possible given the size of the MSR innermost layer. However, the solenoid's mechanical support requires a gap between the coil and the MSR walls. This gap weakens the benefit of the coupling between the shield and the coil and decreases the field uniformity. A remediation was achieved by adding seven end-cap loops located at both coil extremities on the top (bottom) horizontal planes. In summary, the $B_0$ coil is made up of two components connected in series: a square vertical solenoid and two sets of seven end-caps loops.  

The design of the $B_0$ coil was performed with a finite element method simulation (COMSOL software). The goal of the simulation was twofold: define the detailed coil geometry which provides a $B_0$ field uniformity meeting the requirements and estimate the amplitude of the remaining field non\hyphen  uniformities. The simulation included only the innermost MSR layer, as the addition of a second layer has a negligible influence on the generated magnetic field. This layer was defined as a cube with an inner side length of $293~\unit{cm}$ and a thickness of $6~\unit{mm}$ (as well as a few additional bands with a thickness of $7.5 ~\unit{mm}$ used to reinforce the wall structure in the experiment \cite{MSR}). The relative permeability of the wall material was set to $\mu = 35000$. All openings were taken into account. The symmetries of the $B_0$ coil allowed the simulation of only one eighth of the system volume, defined by the following conditions on the coordinates: $x>0$, $y>0$, and $z>0$. The boundary conditions were defined as follows. Outside the MSR the magnetic field is zero at a large distance. Inside the MSR, the symmetry planes are defined as magnetic insulation boundary for the vertical planes XZ and YZ (these planes are anti-symmetric for the coil currents) and as perfect magnetic conductor boundary conditions for the horizontal plane XY (this plane is symmetric for the coil currents).

The solenoid characteristics are constrained by the experimental environment: the solenoid length ($273 ~\unit{cm}$) is limited by the MSR height and the volume required for its support. Similarly, the vertical gap between two adjacent loops, $d_z = 15 ~\unit{mm}$, is set to a minimal value, offering at the same time a sufficient density of surface current (for the production of a uniform field) and a gap between two loops large enough for the coil to be attached to its mechanical support. As a result, the optimization procedure is mainly carried out by varying the number and the shape of the end-caps loops. The variable used for the minimization is the transverse magnetic field $B_T = \sqrt{B_x^2 + B_y^2}$ calculated in the MSR central volume ($ 1~ \unit{m^3}$). 

\begin{figure}[ht!]
  \centering
  \centering \includegraphics[trim = 5.5cm 1cm 0.5cm 1.5cm, clip, scale=0.33]{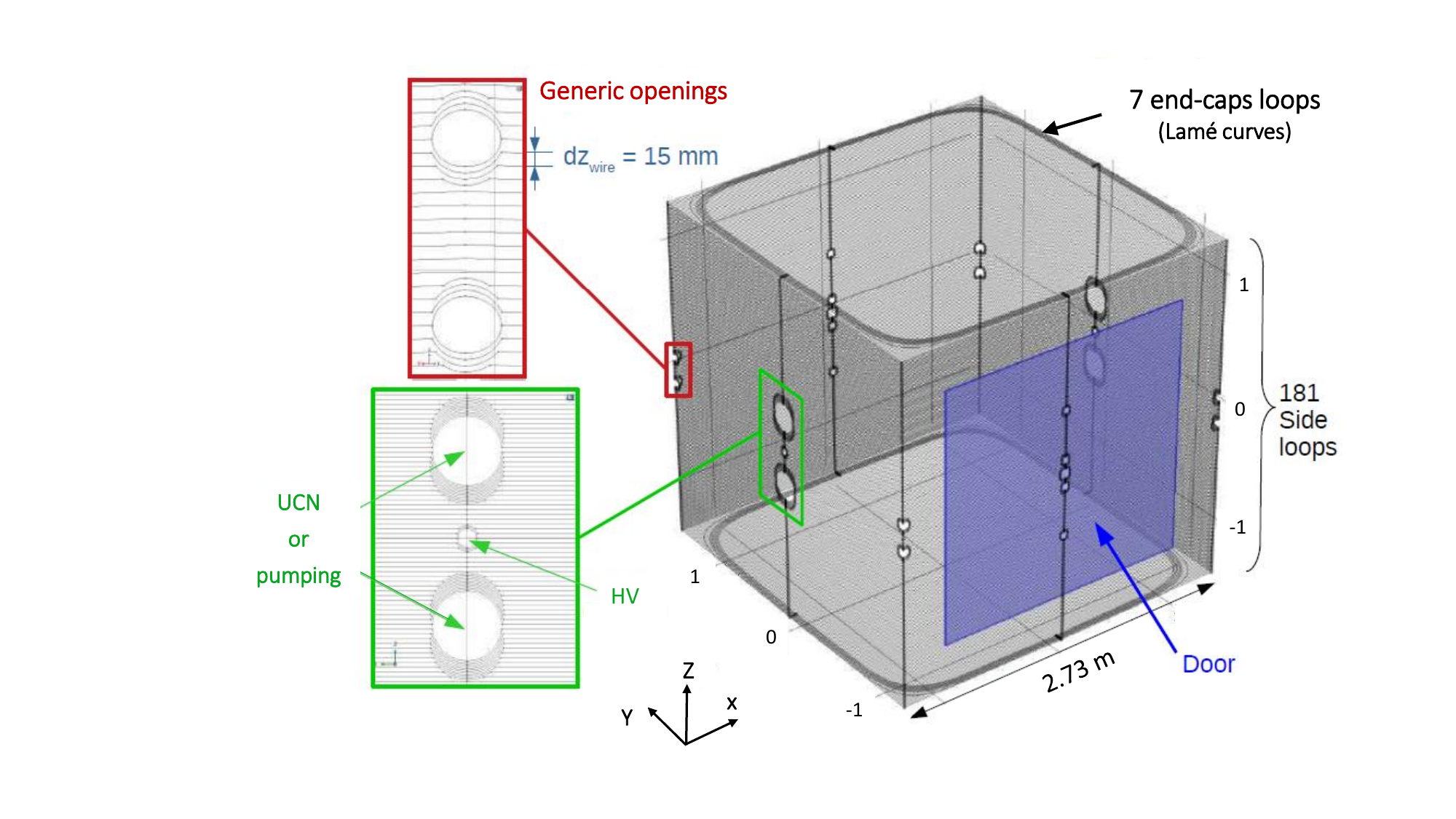}
  \caption{Design of the $B_0$ coil. The Lamé curves are located at the solenoid extremities in the top and bottom horizontal planes. The red and green frames show a detailed view of the opening bypasses. The inner volume of the coil is accessed through a square door (drawn in blue) with a side length of $200 ~\unit{cm}$.}
  \label{fig_B0_coil_design}
\end{figure}

The optimized geometry of the $B_0$ coil is a square solenoid attached to a cubic support fixed at about $10 ~ \unit{cm}$ from the innermost layer of the MSR. The solenoid is made of 181 loops vertically spaced by $15~ \unit{mm}$, complemented by two identical sets of 7 end-cap loops. Their design is parametrized by the Lamé curves, which is an interpolation between a square and a circle:   
\begin{equation}
\left\{
\begin{array}{ll}       
     x_i = a_i  \cos^{n_i}(\varphi) \\
     y_i = a_i \sin^{n_i}(\varphi) \\
     z_i = \pm 1365~ \unit{mm} \\
\end{array} 
\right.,
\label{Lame}
\end{equation}
where $x$ and $y$ are the space coordinates in the horizontal plane (figure \ref{fig_B0_coil_design}), $\varphi$ is the polar angle ranging from $0$ to $\frac{\pi}{2}$ (the full loops are then built by symmetry), and $a_i$ and $n_i$ are the parameters of the Lamé curves $i$ with $i \in [1,7]$. 
For all curves, $n_i = 0.30$, and $a_i$ ranges from $1300 ~\unit{mm}$ to  $1360 ~\unit{mm}$ with a $10 ~\unit{mm}$ step.

The most important deviations from the ideal solenoid result from its various openings. In order of importance, this concerns the openings for the UCN guides ($\times 2$) and the vacuum pipes ($\times 2$) with a diameter of $220 ~\unit{mm}$, the openings for the laser beam used for the Hg co-magnetometer, the high voltage feed through, and other miscellaneous holes with diameters ranging from  $55~ \unit{mm}$ to $160~ \unit{mm}$ (see \cite{MSR} for more details). The gap between loops which bypass the openings is reduced in order to compensate for the lack of loops at the opening location. Two examples are shown in the red and green inserts of figure \ref{fig_B0_coil_design}. In one side of the solenoid, a $\qty{2}{m} \times \qty{2}{m}$ door, depicted as a blue parallelogram, gives access and permits the transit of experimental components. The door panel is equipped with wires closing the solenoid loops. The electrical continuity between wires of the outer walls and of the door panel is ensured by 133 custom-made non-magnetic connectors. Their design permits a current path barely deviating from a straight direction. The entire door can be removed for the insertion of large components such as the precession chambers. 

Magnetic characteristics of all material or pieces used for the construction of the coils were measured before installation on site. Small pieces were checked at PSI \cite{Kletzl2025} while large ones were measured inside the Berlin magnetically shielded Room-2 at Physikalisch-Technische Bundesanstalt \cite{Bork2000}. Weakly magnetizable material were selected: polycarbonate (coil support plates), Aluminum (coil structure), Copper (wire), polylactic acid (3D printed door connectors) and titanium and polyamide (screws). The limit set to select a material or a piece was a maximum magnetic field of $\qty{200}{pT}$ at $\qty{5}{cm}$ distance after exposing the surface to a magnetic field of about $\qty{30}{mT}$. Some magnetic contamination was detected, primarily on the surface of the machined pieces. To address this, specific cleaning procedures were applied, including baths with an alcaline detergent and/or an acidic solution.
Bulk contamination was also identified in a few cases, such as with screws. Approximately 10\% of the titanium screw batches exhibited contamination. In such instances, the affected screws were replaced.


\begin{figure*}[ht!]
  \centering
    \includegraphics[width=0.4\textwidth]{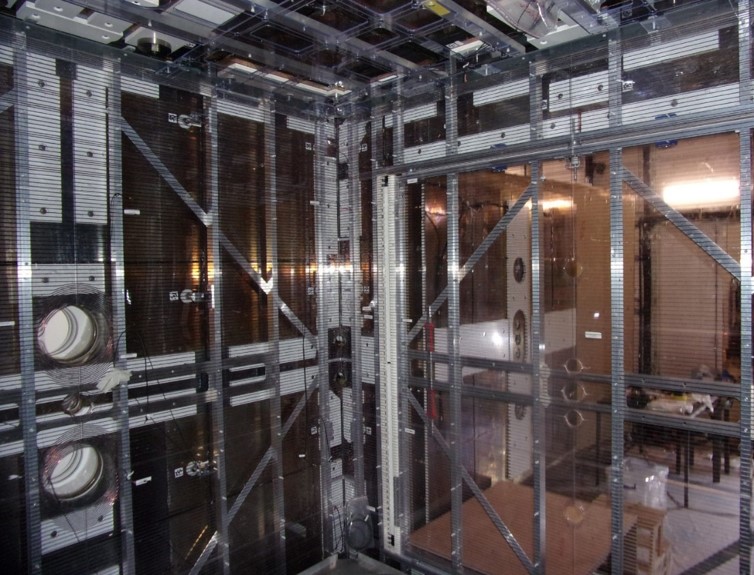}
    \includegraphics[width=0.31\textwidth]{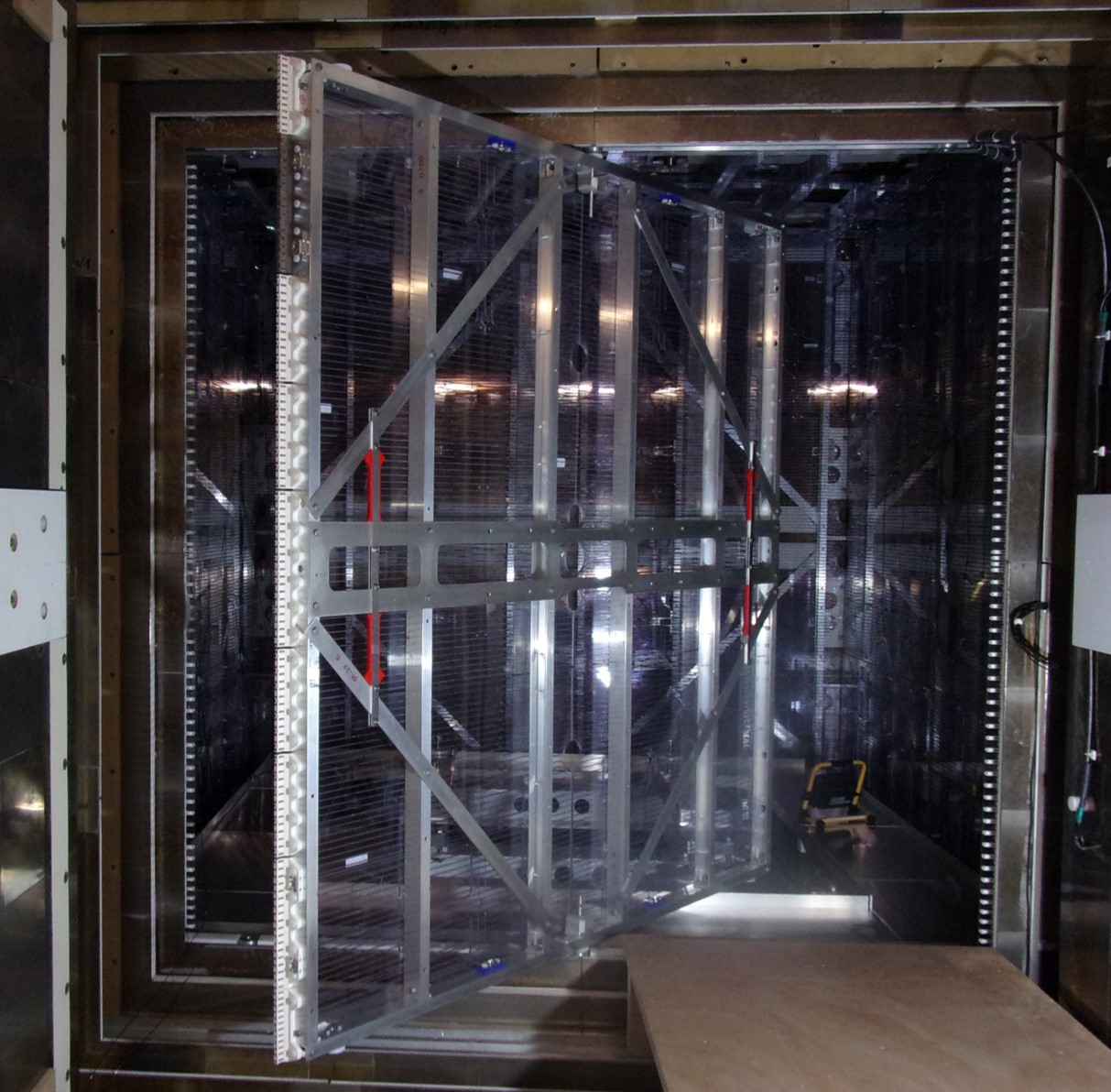}
    \includegraphics[width=0.153\textwidth]{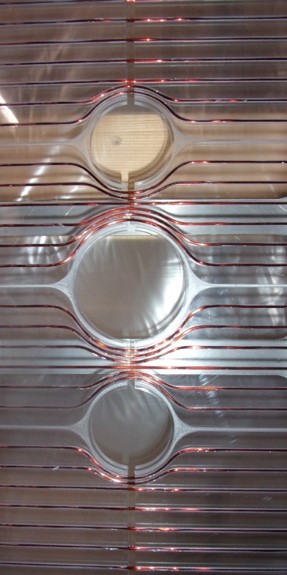}
  \caption{Pictures of the built coil system. Left: inside of the $B_0$ coil. The two openings for the vacuum pipes (UCN guides on the other side) are visible on the left and the $B_0$ door is closed. Middle: the $B_0$ door open. On the door edges, the white custom-made connectors are visible. Right: a typical wire bypass of openings in the $B_0$ coil.}
  \label{B0_photo}
\end{figure*}

The $B_0$ coil wire has a diameter of $\qty{1.8}{mm}$ and reaches a length of $2121~\unit{m}$. Its calculated resistance is $11.3~\unit{\Omega}$ (the addition of the connectors' resistance leads to a measured resistance of $20 ~\unit{\Omega}$). The coil constant, extracted from COMSOL simulations performed with the MSR, is equal to 84 nT/$\unit{mA}$. The field non\hyphen uniformity computed inside each precession chamber is $\sigma(B_z) = 13~\unit{pT}$. The magnetization of the innermost MSR layer plays a crucial role. It contributes to approximately one third of the $G_{0\,0}$ term, the uniform component of the magnetic field in the $z$-direction, and improves by more than two orders of magnitude the field uniformity. However, in the experiment, the coil will be likely not perfectly symmetrical due to unavoidable mechanical imperfections. Their influence needs to be studied in order to establish robust conclusions about field uniformity. 


\subsection{Characterization of field non\hyphen  uniformities} \label{sec_coil-symmetries}

The description of the field non\hyphen  uniformities can be split in two parts: the field non\hyphen  uniformities produced by the designed $B_0$ coil (i.e. a mechanically perfect coil) and the ones resulting from mechanical imperfections.

The symmetries of the designed $B_0$ coil (here the $B_0$ coil term refers to the coil itself and the innermost layer of the shield) propagate to its generated magnetic field, which can only consist of a series of ``allowed'' harmonic modes. A perfectly symmetric $B_0$ coil produces a low number of gradients while any symmetry-breaking allows other gradients to exist (see \ref{sec_symmetry}). Therefore, the system must be as symmetric as possible. All openings in the MSR and/or the $B_0$ coil are symmetrically mirrored on opposite walls as shown in figure \ref{fig_B0_coil_design}, hence preserving the reflection symmetries w.r.t. the XY, YZ and XZ planes. The latter can however be broken by imperfect features of the MSR and $B_0$ doors. In this spirit, the next paragraph lists the number of modes produced by the designed $B_0$ coil. 

The allowed modes generated by a finite solenoid are given by the $\Pi_{2k,4n}$ terms, where $k,n\in\mathbb{N}$. They include the uniform vertical component of the $B_0$ field, corresponding to the mode $\Pi_{0\,0}$, and non\hyphen  uniform modes $\Pi_{2\,0}, \Pi_{4\,0}, \Pi_{4\,4}, \hdots $. The magnitude, $G_{l\,m}$, of the different modes is usually decreasing with mode degree $l$, meaning that $G_{2\,0}$ is the predominant gradient. The presence of the openings breaks the $R_z$ symmetry ($\pi/2$ rotation around the vertical axis). This extends the set of allowed modes to all $\Pi_{2k,2n}$, $k,n\in\mathbb{N}$, the dominant one among the newly allowed modes being $\Pi_{2\,2}$. They lead to the non\hyphen  uniformity of $\sigma(B_z) = 13~\unit{pT}$ stated in the previous section. 

The presence of mechanical imperfections can strongly alter this picture. 
A vertical displacement of the entire $B_0$ coil with respect to the MSR is the main matter of concern. Such a misalignment breaks the reflection symmetry w.r.t. the XY plane, allowing the existence of $\Pi_{2k+1,2n}$ modes ($k,n\in\mathbb{N}$). More precisely, a vertical shift of the $B_0$ coil produces a $G_{1\,0}$ gradient (which can easily exceed the limit defined by the top-bottom matching condition \eqref{eq_gtb}), as well as higher $l$-odd, $m=0$ gradients $\Acute{G}_{3\,0}, \Acute{G}_{5\,0} ...$, inducing a frequency shift mimicking an EDM signal. The extent of this issue was investigated using simulations of the $B_0$ coil placed at different heights with respect to the MSR. Vertical displacements, $\delta z$, between the two systems ranging from 0 to $5 ~\unit{mm}$ were considered. The $G_{1\,0}$ sensitivity to the displacement $\delta z$  derived from this set of simulations is $G_{1\,0}/\delta z  = 6.45 ~ (\unit{pT/cm})/\unit{mm}$, meaning that a $0.1 ~\unit{mm}$ displacement already exceeds the top-bottom matching condition. Dedicated coils, described in section \ref{sub_section_suppression}, were designed to compensate the gradients induced by tiny vertical misalignment. The production of higher $l$-odd, $m=0$ gradients was also observed. Sensitivities to the displacement $\delta z$ are reported for normalized gradients in table \ref{Gradient_displacement_sensitivity}. While the $\Acute{G}_{ 5\,0}/\delta z$ and $\Acute{G}_{ 7\,0}/\delta z$ sensitivities are weak and can be accommodated, the $\Acute{G}_{3\,0}/\delta z$ sensitivity is substantial and requires specific care. In order to address this flaw, the height of the $B_0$ coil support was made adjustable in a $\pm ~3~\unit{mm}$ range. In the case of a misalignment, one can change the height of the $B_0$ coil and determine the optimal vertical position by measuring the vertical gradient (see section \ref{sec_characterization}). 


\begin{table}[htb!]
\centering
\begin{tabular}{@{} c c @{}}
     \toprule
     Sensitivity  & Values (fT/cm)/mm  \\
     \midrule
      $G_{1\,0}/\delta z$ & 6450   \\
      $\Acute{G}_{3\,0}/\delta z$ & 38   \\
      $\Acute{G}_{ 5\,0}/\delta z$ & 6.1 \\
      $\Acute{G}_{7\,0}/\delta z$ & $7.3\times10^{-2}$ \\
      \midrule
      $G_{1\,1}/\delta x$ & 360 \\
      $G_{1\,-1}/\delta y$ & 340 \\
     \bottomrule
\end{tabular}
\caption{Sensitivities of the $l$-odd, $m=0$ normalized gradients to the vertical displacement $\delta z$ and sensitivities
of the $G_{1\,1}$ and $G_{1\,-1}$ gradients to the horizontal displacements $\delta x$ and $\delta y$.}
\label{Gradient_displacement_sensitivity}
\end{table}

Horizontal displacements of the $B_0$ coil along the X and Y directions are less penalizing. They break the reflection symmetry w.r.t. the XZ and YZ planes and respectively allow $\Pi_{2k+1,-2n-1}$ and $\Pi_{2k+1,2n+1}$ modes, with $k,n\in\mathbb{N}$. The new possible gradients alter the field uniformity to a limited extent, increasing $\sigma(B_z)$ by a few $\unit{pT}$ for displacements of $5 ~\unit{mm}$ in both horizontal directions.


The MSR layers are made of several mu-metal plates between which the relative permeability may vary by at most 20 $\%$ \cite{MSR}. Such a variation between the roof and the floor layers breaks the $z$-symmetry and may introduce a source of non\hyphen uniformity for the vertical magnetic field component. A simulation with a difference of 20 $\%$ between the permeability of the roof and the floor material showed no relevant decrease of magnetic field uniformity. We conclude, that the material's absolute permeability is large enough making 20$\%$ relative changes negligible.

Possible displacements of the $B_0$ coil wire from its ideal path may also be a source of non\hyphen uniformity. Taking into account the groove width in which the $B_0$ wire is inserted, $\qty{2}{mm}$, and the wire diameter, $\qty{1.8}{mm}$, simulations were performed with undulating wires (with a periodicity and a phase at the origin different from one loop to another). They did not show any significant influence on the field uniformity, likely due to an overall compensation effect between all loops. 

Finally, a more realistic model of the MSR innermost layer is implemented in the COMSOL simulation after construction of the MSR. This simulation takes into account the exact wall dimensions, slightly larger than in the ideal model ($\Delta x = +2.3 ~\unit{mm}$, $\Delta y = +2.8 ~\unit{mm}$, $\Delta z = +0.6 ~\unit{mm}$). Furthermore, the coil geometry now includes a recession of the MSR door ($\Delta y = 6 ~\unit{mm}$), a feature that breaks the $\sigma_y$ symmetry (reflection in the XZ plane). This allows the $\Pi_{2k+1,-2n-1}$, $k,n\in\mathbb{N}$, modes in the coil's harmonic spectrum, the dominant one being $\Pi_{1\,-1}$, on top of the already allowed $\Pi_{2k,2n}$, $k,n\in\mathbb{N}$, modes. All of these allowed gradients are produced in the simulation and recorded in table \ref{tab_spectrum}. The field non\hyphen uniformity in each precession chamber $\sigma(B_z)$ resulting from this model is increased from $13 ~\unit{pT}$ to $16~\unit{pT}$.

\subsection{Design of the auxiliary coils}
\label{sub_section_suppression}

The remaining magnetic field non\hyphen uniformities can be suppressed by adjusting currents in the correction coil array. The array is made of 9 independent ($30 \times 30~\unit{cm}$) square coils mounted on each surface of the $B_0$ coil support (there are indeed 10 coils on the sides where the UCN guide and the vacuum pipe openings are located). A full description of the coils is given in figure 20 of \cite{n2edm}. The array can produce harmonic modes $\Pi_{lm}$ up to the $6^{th}$ $l$-order and therefore can be used to suppress all harmonics at orders lower than the $6^{th}$. The procedure used to optimize the field uniformity is described in section \ref{sec_optimization}.

Seven additional coils, the gradient coils, can also be used to produce specific gradients of 
the $B_z$ components \cite{n2edm}. The $G_{1\,0}$, $G_{1\,1}$ and $G_{1\,-1}$ coils produce the linear gradients $\partial_zB_z$, $\partial_xB_z$ and $\partial_yB_z$ while the $G_{2\,0}$ and $G_{3\,0}$ coils describe the quadratic and cubic gradients of the $B_z$ component. Finally, the $G_{0\,1}$ and $G_{0\,-1}$ produce the constant term of the horizontal components $B_x$ and $B_y$ respectively. Beside their role in the optimization 
of the magnetic field, they are used to fulfill the top-bottom matching condition ($G_{1\,0}$), to control the gradients that induce a false motional EDM ($G_{1\,0}$ and $G_{3\,0}$), to achieve the requested field uniformity ($G_{2\,0}$), and to benchmark the Cs magnetometer locations ($G_{1\,1}$ and $G_{1\,-1}$). Their geometrical description is given in \ref{Gradient_coils}.

All coils are powered by true bipolar current sources developed in the collaboration. The power supplies of the optimization coils have a current range of $\pm~200~\unit{mA}$ with a setting resolution of \SI{1}{\micro A} while the current ranges for the gradient power supplies is $\pm~20~\unit{mA}$ with a setting resolution of \SI{0.1}{\micro A}. The current stability was assessed by the Allan standard deviation measured with an applied current of 10~mA or 100~mA. The deviation was found to be below 1 ppm after 3 min which fulfills the requirements defined in \cite{n2edm}. 

\section{Characterization of the \texorpdfstring{$B_0$}{B0} coil} \label{sec_characterization}

\subsection{Magnetic field mapping}

\begin{figure}[ht!]
    \centering
    \includegraphics[width=0.49\textwidth]{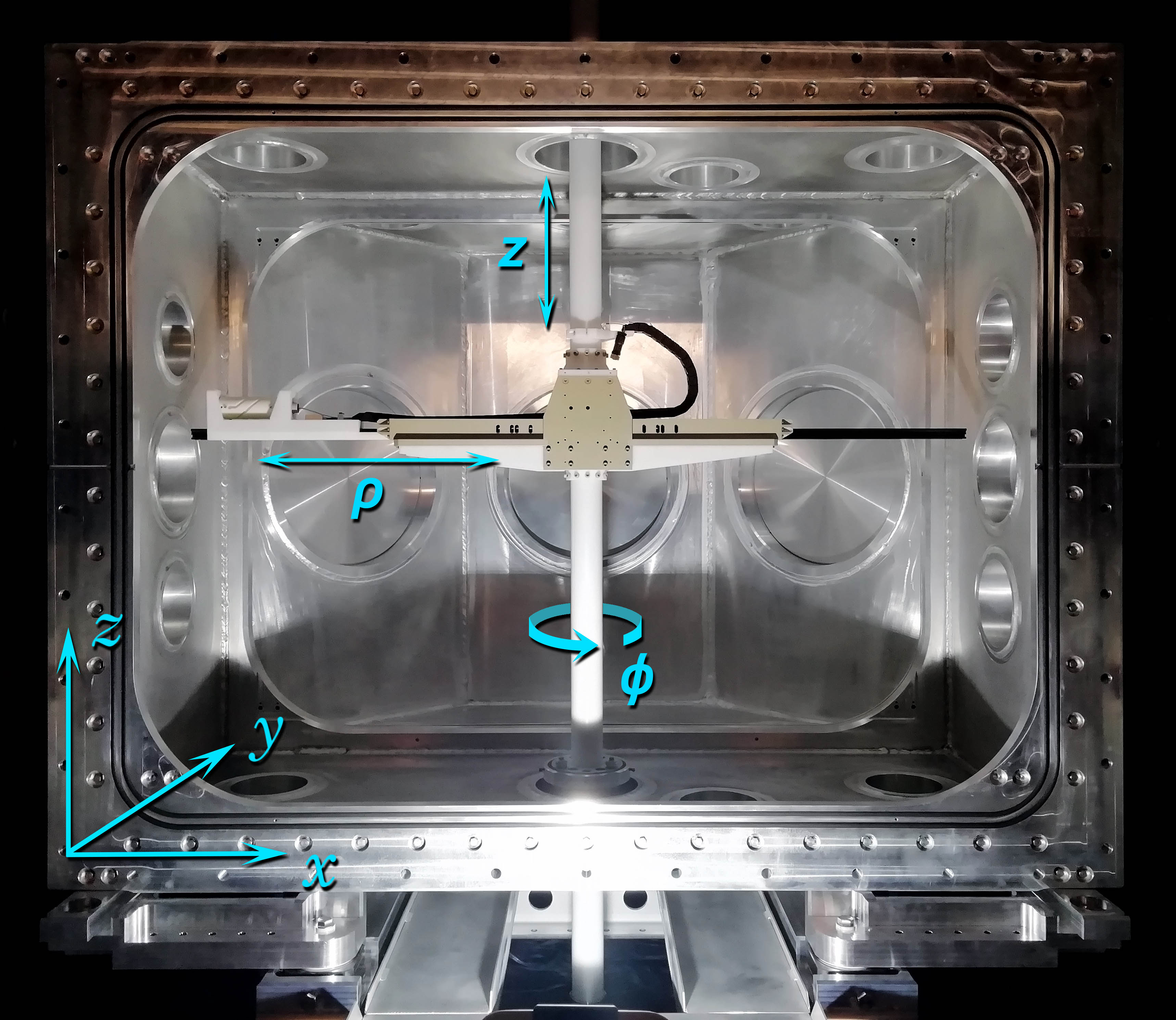}
    \caption{The n2EDM mapper inside the empty vacuum vessel. The fluxgate mounted on the mapper's arm can travel to any point inside a cylindrical volume of radius $\qty{78}{cm}$ and height $\qty{82}{cm}$, at which it measures the three projections of the magnetic field.}
    \label{fig_mapper}
\end{figure}

\begin{figure}[ht!]
    \centering
    \includegraphics[width=0.49\textwidth]{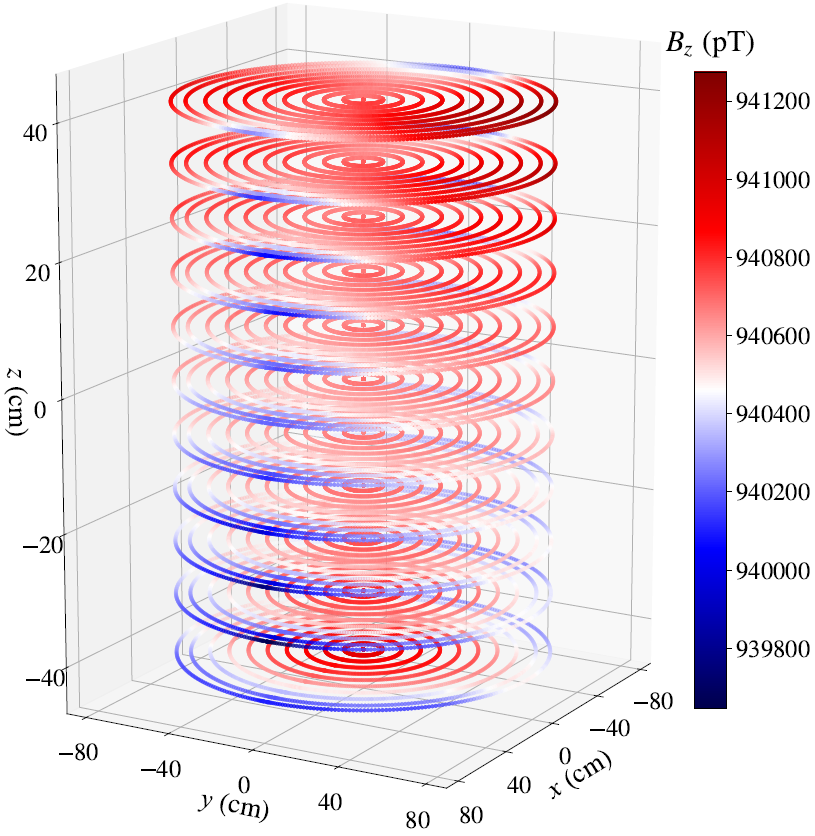}
    \caption{An example of a magnetic field map of the field generated by the n2EDM coil system and recorded by the mapper. Each point corresponds to the vertical projection of the magnetic field inside a cylindrical volume of radius $\qty{78}{cm}$ and height $\qty{82}{cm}$.}
    \label{fig_map}
\end{figure}

The offline mapping of the magnetic field is performed using an automated field mapper, pictured in figure \ref{fig_mapper}. The mapper consists of a three-axis low-noise Bartington MAG13 fluxgate \cite{bartington}, mounted on a motorized arm that allows it to explore a cylindrical volume of $\qty{78}{cm}$ radius and $\qty{82}{cm}$ height. The fluxgate can also be rotated along the $\rho$ axis to determine its DC-offset. Magnetic field maps are recorded in a series of rings $(\rho, z)$ performed inside a given cylindrical volume. One ring takes $\qty{74}{s}$, during which the fluxgate measures the field with a sampling frequency of $\qty{10}{Hz}$. A plot of the vertical field projection of a $B_0$ coil map is shown in figure \ref{fig_map}.

The characterization of the magnetic field consists in the extraction of the harmonic spectrum $\{G_{lm}\}_{l\leq 7}$, with $-l-1\leq m \leq l+1$, of the polynomial expansion \eqref{eq_harmonic-expansion} from a map. This extraction is performed by first fitting all rings $(\rho, z)$ with a Fourier series in $\varphi$, and then fitting the Fourier coefficients with the polynomials functions in $\rho$ and $z$ of the harmonic expansion. The fits presented here are done up to order $l=7$. This procedure was already employed for the nEDM experiment and is explained more thoroughly in \cite{mapping2022}. While we present here the most significant measurements of the n2EDM mapping campaign, the full results are discussed in chapters 7, 8, and 9 of \cite{bouillaud}.

The following subsections will demonstrate that all requirements on magnetic field uniformity are satisfied by analyzing magnetic field-mapping data. All the measured $B_0$ field maps presented here were recorded with a coil current $I=\qty{11.25}{mA}$.

\subsection{Magnetic field uniformity}

\begin{figure*}[ht!]
    \hspace{-0cm}\centerline{
    \includegraphics[width=0.31\textwidth]{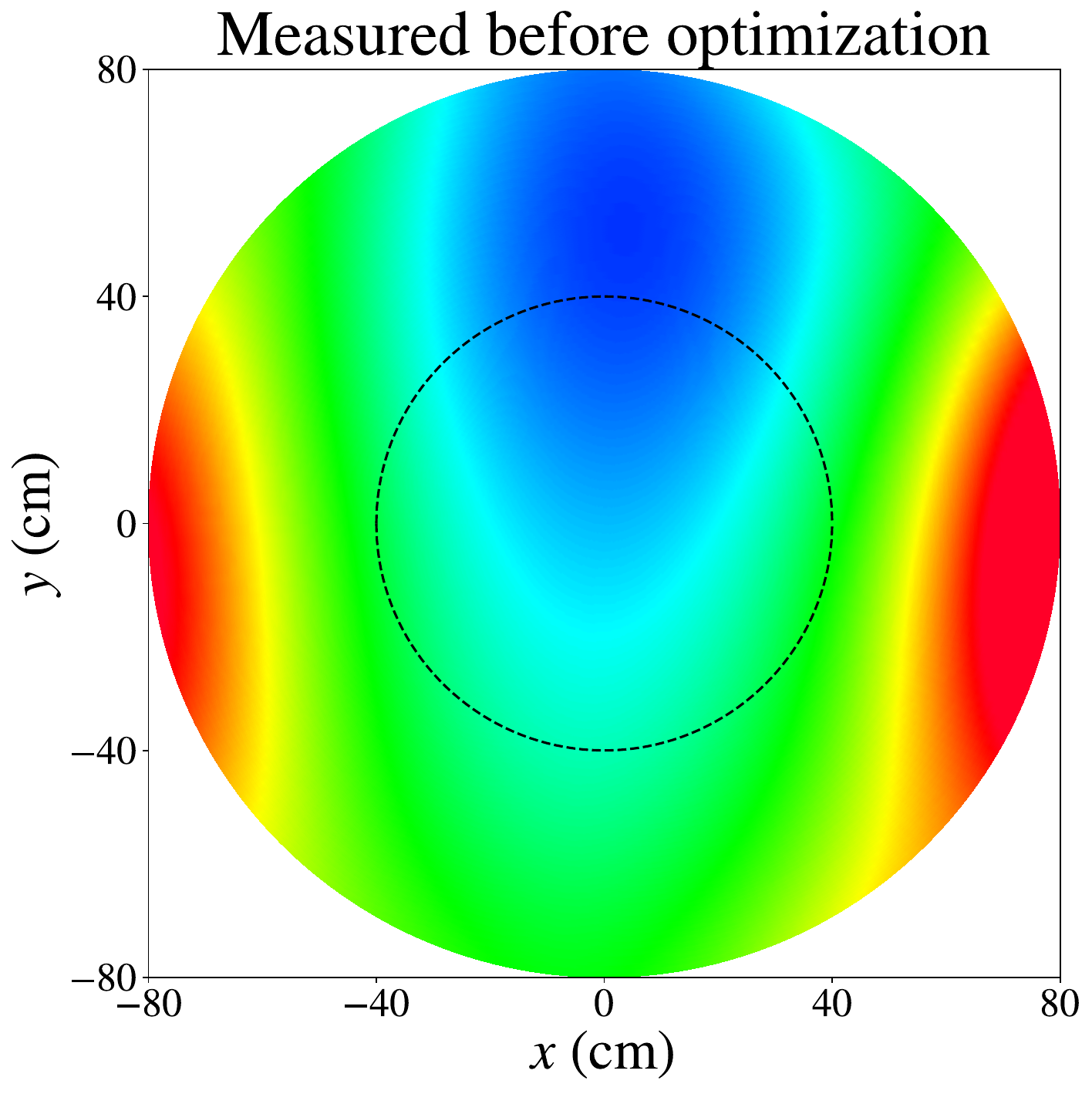}
    \includegraphics[width=0.31\textwidth]{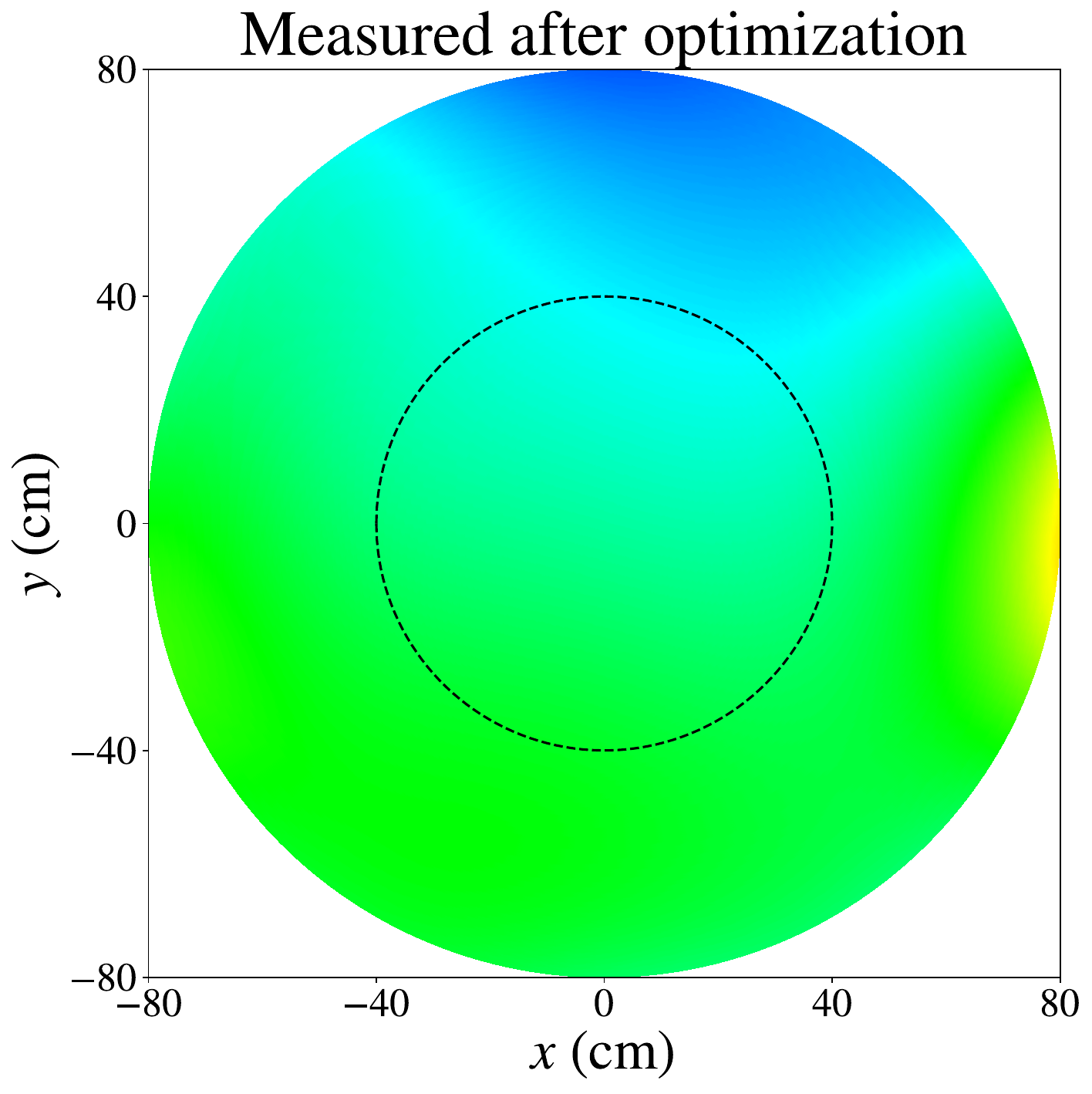}
    \includegraphics[width=0.38\textwidth]{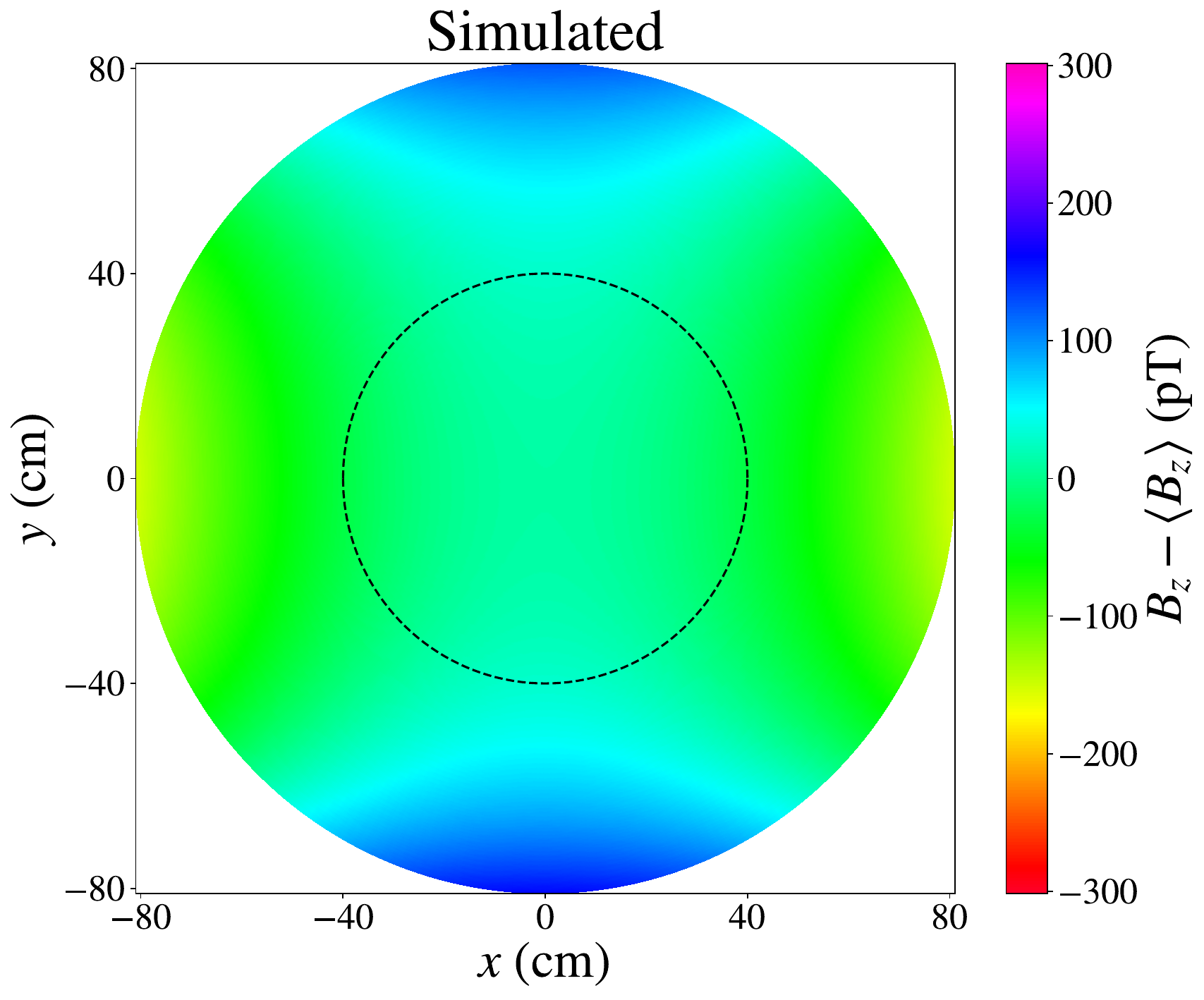}
    }
    \caption{Horizontal cut at $z=0$ of the deviations of the vertical $B_0$ magnetic field, in the positive coil polarity. The simulated values from \cite{flaux} are compared to the 2022 measurements before and after optimization with auxiliary coils. The latter successfully cancels the main contributions of the $\boldsymbol{\Pi}_{2\,0}$ and $\boldsymbol{\Pi}_{2\,2}$ modes. The dashed black lines show where the walls of the precession chambers are positioned.}
    \label{fig_uniformity}
\end{figure*}

\begin{figure}[ht!]
    \centering
    \includegraphics[width=0.49\textwidth]{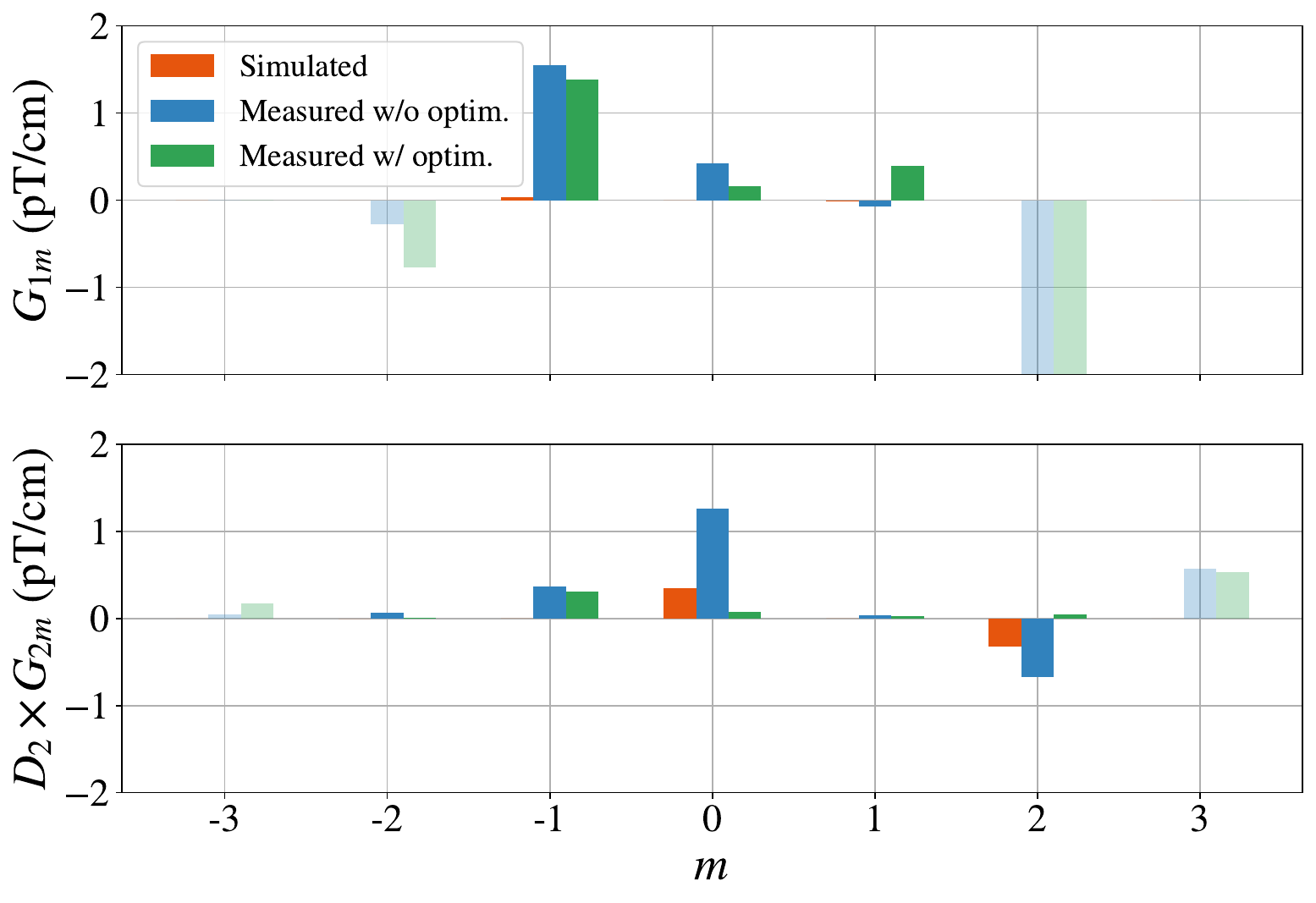}
    \caption{Harmonic spectrum of the $B_0$ coil, in the positive polarity, for the simulated field and measured field before and after optimization. The fit is performed up to order $l=7$ but here only modes of indices $l=1,2$ are shown. The considered volume for all spectra is the mapped cylindrical volume. Note that purely transverse harmonic modes with $m=\pm(l+1)$ (in faded colors) do not contribute to the total non\hyphen  uniformity $\sigma(B_z)$.}
    \label{fig_spectrum}
\end{figure}

The vertical component of the simulated and measured $B_0$ fields, in the $z=0$ plane, is plotted in figure \ref{fig_uniformity}. Moreover, figure \ref{fig_spectrum} shows the corresponding harmonic spectra. Even though the measured field is not as uniform as the simulated field, it nevertheless satisfies the requirement \eqref{eq_uniformity}.
As discussed earlier, the relevant quantity for n2EDM is the RMS on the vertical field component inside each precession chamber, which for the measured $B_0$ field amounts to
\begin{align}
    \sigma(B_z)_\text{TOP} &= \qty{48}{pT}, & \sigma(B_z)_\text{BOT} &= \qty{38}{pT}.  \label{eq_uniformity-num}
\end{align}
For comparison, the residual field in the MSR without the $B_0$ coil has a uniformity of $\sigma(B_z)_\text{TOP}=\qty{15}{pT}$ and $\sigma(B_z)_\text{BOT}=\qty{11}{pT}$. The $B_0$ coil is then the primary source of non\hyphen  uniformities. 

\sloppy The vertical field RMS of equation \eqref{eq_uniformity-num} receives contributions from all harmonic modes. The orthogonality of the trigonometric functions in $m\varphi$ enforces that harmonic modes of different $m$ index add up quadratically inside $\sigma(B_z) = \sqrt{\left<(B_z-\left<B_z\right>)^2\right>}$, while those with same index $m$ can interfere. As apparent in figure \ref{fig_spectrum}, the dominant contributions are $|G_{1\,-1}|\sigma(\Pi_{z,1-1}) = \qty{31}{pT}$, $|G_{2\,0}|\sigma(\Pi_{z,20}) = \qty{17}{pT}$, and $|G_{2\,2}|\sigma(\Pi_{z,22}) = \qty{24}{pT}$. The quadratic sum of these three contributions amounts to $\qty{43}{pT}$. The interference between the $\Pi_{z,20}$ and $\Pi_{z,30}$ modes is responsible for the difference between $\sigma(B_z)_\text{TOP} $ and $\sigma(B_z)_\text{BOT} $.

The horizontal profile of the vertical field component depicted in figure \ref{fig_uniformity} is also well described by its dominant harmonic modes. 
Considering only the $\Pi_{2\,0}$ and $\Pi_{2\,2}$ contributions from the polynomial expansion \eqref{eq_harmonic-expansion}, we write the vertical field component as $B_z = -(G_{2\,0}/2 - G_{2\,2})x^2 - (G_{2\,0}/2 + G_{2\,2})y^2$. The parabolic shape witnessed in the horizontal field profile is consistent with this expression when plugging in the generalized gradients from figure \ref{fig_spectrum}. Visually, it can be thought as a linear combination of the graphical representations of $\boldsymbol{\Pi}_{2\,0}$ and $\boldsymbol{\Pi}_{2\,2}$ included in \ref{fig_harmonic-z0}. 
As for the $y$-odd structure that appears only in the measured fields, this can be attributed to the presence of $l$-odd harmonic modes, especially of $\boldsymbol{\Pi}_{1\,-1}$. The vertical field consisting only of this mode is written $B_z=G_{1\,-1} y$. The sign of $G_{1\,-1}$, which is non-zero because of the presence of the MSR door, explains the global shift of the parabola in the transverse plane.


Overall, the magnetic field generated by the $B_0$ coil achieves the desired uniformity consistent with a departure from the ideal coil symmetry taking into account the presence of the neutron guides, vacuum pipes, and MSR door. This is shown by the measured harmonic gradients of table \ref{tab_spectrum}. Furthermore, its vertical component satisfies the n2EDM uniformity requirements.


\begin{table}[ht!]
\begin{tabular}{@{} l c c c @{}}
    \toprule 
    \multicolumn{4}{l}{\textbf{Allowed by idealized symmetry}} \\
    Gradients & $\Acute{G}_{2\,0} ~ \unit{(pT/cm)}$ & $\Acute{G}_{4\,0}~\unit{(pT/cm)}$ & $\Acute{G}_{4\,4}~\unit{(pT/cm)}$ \\
    \midrule
    Simulated & $0.32$ & $\num{6.90e-2}$ & $\num{-0.94e-3}$ \\ 
    Measured & $1.27$ & $\num{-12.25e-2}$ & $\num{-12.20e-3}$ \\
    \midrule    
    \multicolumn{4}{l}{\textbf{Allowed by hole-broken symmetry}} \\
    Gradients & $\Acute{G}_{2\,2} ~ \unit{(pT/cm)}$ & $\Acute{G}_{4\,2}~\unit{(pT/cm)}$ & \\
    \midrule
    Simulated & $-0.30$ & $\num{5.91e-2}$ & \\ 
    Measured & $-0.67$ & $\num{-1.44e-2}$ & \\ 
    \midrule
    \multicolumn{4}{l}{\textbf{Allowed by door-broken symmetry}} \\
    Gradients & $\Acute{G}_{1\,-1} ~ \unit{(pT/cm)}$ & $\Acute{G}_{3\,-1}~\unit{(pT/cm)}$ & $\Acute{G}_{3\,-3}~\unit{(pT/cm)}$ \\
    \midrule
    Simulated & $0.04$ & $\num{2.09e-2}$ & $\num{0.47e-2}$ \\ 
    Measured & $1.54$ & $\num{5.18e-2}$ & $\num{-0.56e-2}$ \\ 
    \bottomrule
\end{tabular}
    \caption{Simulated and measured values of the harmonic coefficients allowed by three coil symmetries.}
  \label{tab_spectrum}
\end{table}

\subsection{Magnetic field gradient generated by a vertical coil displacement}
\label{sec_meas_coil_displac}

\begin{figure}[ht!]
  \centering
  \includegraphics[width=0.49\textwidth]{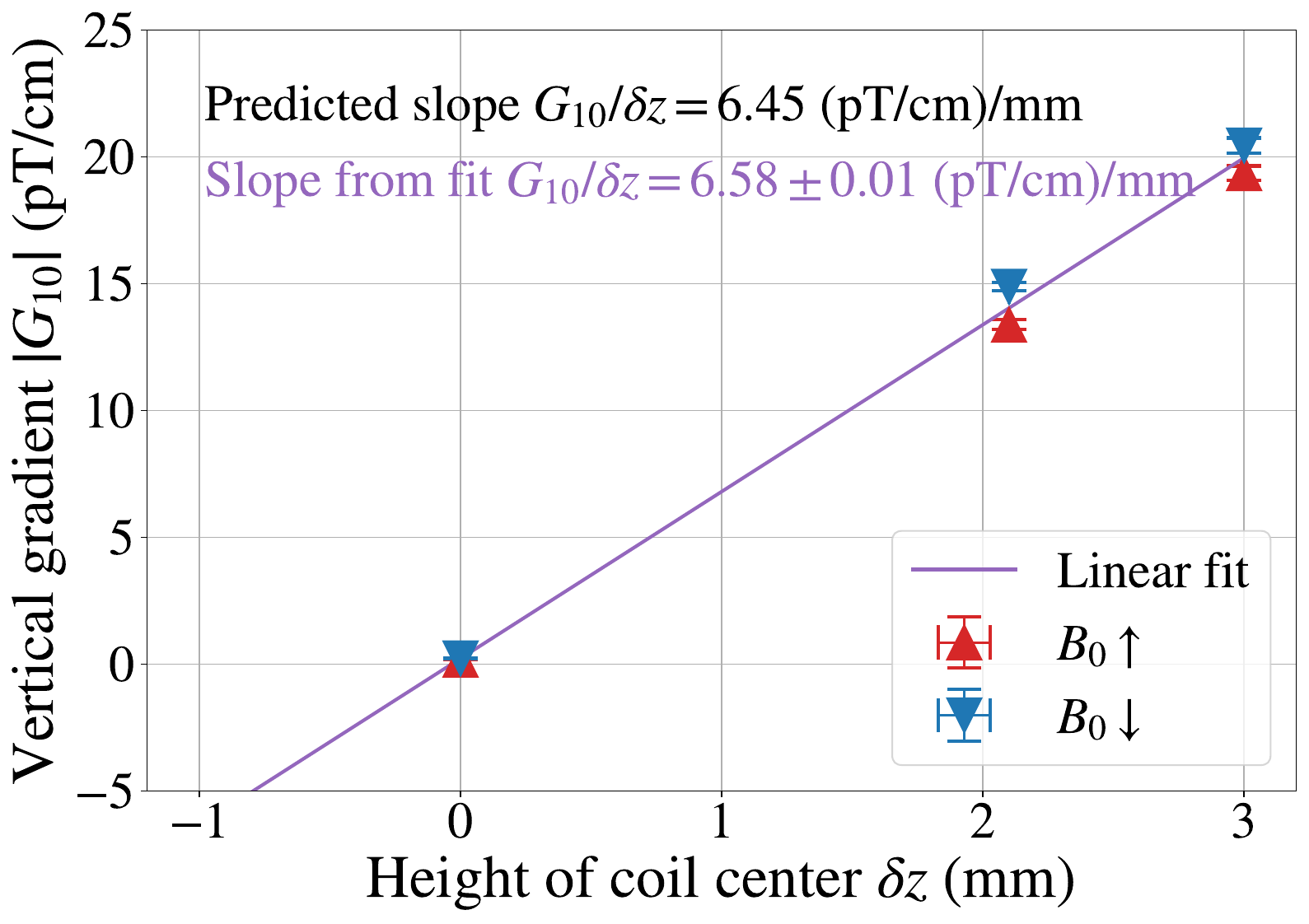}
  \caption{The triangle points show measurements of the vertical magnetic field gradient, in both polarities of the coil, at three different vertical positions of the coil center with respect to the magnetic origin of the MSR. The measurements were taken from right to left in chronological order. The slope of the linear fit matches the predicted gradient value.}
  \label{fig_coilmove}
\end{figure}

Our first measurements of the vertical gradient $G_{1\,0}$ for both $B_0$ coil polarities, taken after mounting the coil and plotted on figure \ref{fig_coilmove} as the rightmost red and blue points, were far above the $\qty{0.6}{pT/cm}$ limit imposed by the top-bottom gradient. As discussed in section \ref{sec_coil-symmetries}, a vertical displacement of the coil with respect to the MSR generates a vertical gradient $G_{1\,0}$ proportional to the displacement, by breaking the reflection symmetry w.r.t. the horizontal plane.
By moving the coil vertically by $\qty{1}{mm}$, we verified that the linear slope matched our calculations. This allowed us to calculate the ideal position, another $\qty{1}{mm}$ lower, satisfying our requirement.
In its final position, the $B_0$ coil satisfies without optimization the top-bottom resonance matching condition \eqref{eq_gtb}.

\subsection{Reproducible gradients for the correction of systematic effects}

We finally determine from the coil's harmonic spectra the problematic phantom modes of orders three, five, and seven generated by the $B_0$ coil alone, and estimate their reproducibility.

To determine the reproducibility of the $B_0$ coil's harmonic spectrum, maps were recorded after a full reset of the experiment's magnetic environment. This reset consists in opening and closing the MSR to allow thermal excitation and relaxation, followed by a procedural demagnetization - or \textit{degaussing} - of the internal field, between each map. We thus define the reproducibility of the magnetic field as the standard deviation of the field over a set of maps separated by such a magnetic reset. Further details on the n2EDM degaussing procedure are given in \cite{degaussing}.

As shown in figure \ref{fig_phantoms}, although the systematic effect generated by modes $\boldsymbol{\Acute{\Pi}}_{3\, 0}$ and $\boldsymbol{\Acute{\Pi}}_{5\, 0}$ of the non-optimized $B_0$ field through equation \eqref{eq_falseEDM-gradients} is above the systematical limit \eqref{eq_falseEDM-condition},
it is in all cases reproducible below this limit.
In other words, the typical variations of the magnetic field are small enough to allow an estimate of the false EDM below the target sensitivity. As mentioned in section \ref{sec_uniformity}, the top-bottom gradient shown in the leftmost set of bars will be accounted for by the online analysis.

While the $B_0$ coil is, as discussed earlier, responsible for the larger share of the non\hyphen  uniformities $G_{lm}$ of the magnetic field, it is however not the limiting factor regarding the reproducibility of the total field.
In fact, the reproducibility measured with and without the $B_0$ coil was found to be of the same level. This is true in particular for the reproducibility of the phantom gradients, shown in figure \ref{fig_phantoms}, which is only slightly larger than without the $B_0$ coil.
A dedicated study showed that the residual field is not a random noise but exhibits a distinctive pattern, as it is generated by thermo-electric currents flowing through the vacuum vessel (chapter 8 of \cite{bouillaud}). In order to maintain this field as stable as possible, the n2EDM MSR is thermally insulated from the rest of the thermohouse and a complete demagnetization of the shield is performed before mapping or data-taking. 


We conclude that the n2EDM magnetic field matches the requirements on field uniformity with respect to both statistical errors and systematical errors. These are summarized in table \ref{table_requirements}. In particular, offline measurements show that problematic phantom modes are reproducible enough to either (A) allow for an estimate of the generated systematic effect through \eqref{eq_falseEDM-gradients}, or (B) successfully implement a field-optimization strategy to cancel these modes and bring the systematic effect below the requirement \eqref{eq_falseEDM-condition}. 

\subsection{Magnetic field optimization with gradient and optimization coils} \label{sec_optimization}

Individual mapping of the 56 optimization coils and 6 gradient coils allow us to determine their respective coil constants. These correspond to the ratio between the current driven through the coil and the amplitude $G_{lm}$ of the generated harmonic modes, for all modes of the spectrum. One can then determine from the measured harmonic spectrum the coil currents that cancel a given set of harmonic modes. 

\begin{figure}[ht!]
  \centering
\includegraphics[width=0.49\textwidth]{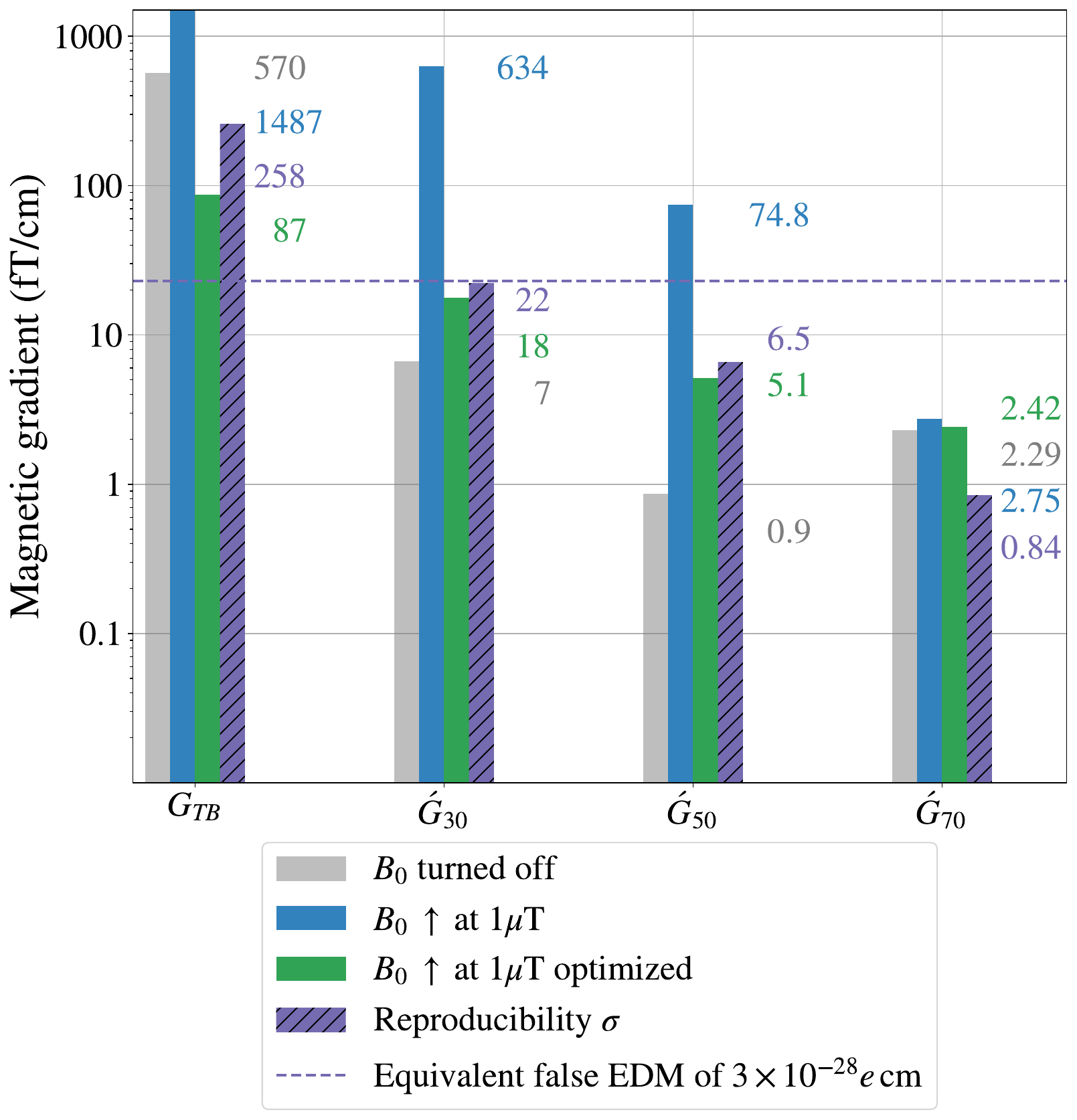}
  \caption{Normalized gradients responsible for the false EDM through equation \eqref{eq_falseEDM-gradients}, extracted from the magnetic field maps. These maps were recorded in three different magnetic configurations: residual field ($B_0$ coil is turned off), $B_0$ coil turned on in the positive polarity, $B_0$ coil turned on in the positive polarity along with a combination of optimization coils and gradient coils that suppress problematic harmonic gradients. The magnetic field reproducibility with respect to a full demagnetization of the MSR is common to non-optimized and optimized $B_0$ measurements. The $\qty{23}{fT/cm}$ limit imposed on the gradients corresponds to a false EDM of $\qty{3e-28}{\ele.cm}$. The effect at order $l=7$ is not significant even without optimization, so the harmonic expansion is not carried out beyond that order.}
  
  \label{fig_phantoms}
\end{figure}

As the primary target of the magnetic field optimization is to suppress the false EDM, we choose a set of currents that cancels harmonic modes $\boldsymbol{\Pi}_{3\,0}$, $\boldsymbol{\Pi}_{5\,0}$, and $\boldsymbol{\Pi}_{7\,0}$. However we are also able to simultaneously cancel other problematic modes $\boldsymbol{\Pi}_{2\,0}$ and $\boldsymbol{\Pi}_{2\,2}$, which, as explained in the previous section, greatly contribute to the non\hyphen uniformity on the vertical magnetic field component. Figure \ref{fig_phantoms} shows that the phantom modes of the optimized $B_0$ field all generate a false EDM below the limit given by equation \eqref{eq_falseEDM-condition}. This agreement is also featured in table \ref{table_requirements}.

Furthermore, the cancellation of $\boldsymbol{\Pi}_{2\,0}$ and $\boldsymbol{\Pi}_{2\,2}$, as visible in figure \ref{fig_spectrum}, reduces the vertical non\hyphen  uniformity in each chamber to
\begin{align}
    \sigma(B_z)_\text{TOP} &= \qty{32}{pT}, & \sigma(B_z)_\text{BOT} &= \qty{21}{pT},  \label{eq_uniformity-opt-num}
\end{align}
nearly one order of magnitude below the statistical requirement. The middle plot of figure \ref{fig_uniformity} confirms that it is indeed the cancellation of the parabolic modes $\boldsymbol{\Pi}_{2\,0}$ and $\boldsymbol{\Pi}_{2\,2}$ that lowers the non\hyphen  uniformity. Finally, the non-uniformity over the volume of interest encompassing the two precession chambers, of radius $\qty{40}{cm}$ and height $\qty{12}{cm}$, amounts to 
\begin{equation}
    \sigma(B_z) = \qty{27}{pT}. \label{eq_uniformity-total-num}
\end{equation}

In conclusion, we are not only able to match the reproducibility requirements for the control of the false EDM, but also to largely cancel the latter. Furthermore, the optimized $B_0$ vertical field is nearly one order of magnitude more uniform than the design requirement.

\begin{table*}[ht!]
\ra{1.2}
\centering
\begin{tabular}{@{\kern\tabcolsep} l c c c @{\kern\tabcolsep}}
    \toprule
    & \textbf{Required} &\textbf{ w/o optim.} & \textbf{w/ optim.} \\
    \textbf{Statistical requirements} & & & \\
    \midrule
    Non-uniformity $\sigma(B_z)_\text{TOP}$ (pT) & $< 170$ & $48\pm 1$ & $32\pm 1$  \\
    Non-uniformity $\sigma(B_z)_\text{BOT}$ (pT) & $< 170$ & $38\pm 1$ & $21\pm 1$  \\
    Top-bottom condition $|G_{1\,0}|$ (pT/cm) & $< 0.6$ & $0.35\pm 0.25$ & $0.16\pm 0.25$ \\
    \midrule
    \textbf{Systematical requirements} & & &\\
    \midrule
    $d_{\text{n}\leftarrow\text{Hg}}^{\text{false}}(\Acute{G}_{3\, 0}\Acute{\Pi}_{3\, 0})$ ($\qty{e-28}{\ele.cm}$) & $ < 3 $ & $81.7 \pm 2.9$ & $2.3 \pm 2.9$  \\
   $d_{\text{n}\leftarrow\text{Hg}}^{\text{false}}(\Acute{G}_{5\, 0}\Acute{\Pi}_{5\, 0})$ ($\qty{e-28}{\ele.cm}$) & $ < 3 $ & $9.2 \pm 0.7$ & $0.7 \pm 0.7$  \\
   $d_{\text{n}\leftarrow\text{Hg}}^{\text{false}}(\Acute{G}_{7\, 0}\Acute{\Pi}_{7\, 0})$ ($\qty{e-28}{\ele.cm}$) & $ < 3 $ & $0.3 \pm 0.1$ & $0.2 \pm 0.1$  \\
    \bottomrule
\end{tabular}
\caption{Table of n2EDM requirements on magnetic field generation for statistical and systematical errors (second column), and their measured values with associated standard deviations (two last columns). The second-to-last column comes from the characterization of the field produced by the $B_0$ coil alone, while the last column concerns the optimized field produced by the $B_0$ coil and a well-chosen combination of optimization coils. The systematical requirements concern both the measured false EDM values and the associated deviations due to field reproducibility.}
\label{table_requirements}
\end{table*}

\section{Conclusion}

We designed and commissioned a coil system to generate a highly uniform magnetic field for n2EDM, an upcoming experiment to measure the electric dipole moment of the neutron with a sensitivity of $\qty{1e-27}{\ele.cm}$.

The design of the coil is a finite square solenoid wired inside a magnetic shield, the MSR of n2EDM \cite{MSR}. The coil together with the shield mimics an infinite solenoid and generates a vertical uniform magnetic field of \SI{1}{\micro T}. 
Furthermore, the solenoid wiring goes around several geometrical features that break the inherent symmetries of a finite square solenoid. 

The field generated by this coil system is expressed in the convenient harmonic polynomial expansion. Because of the conservation of the coil symmetries in the generated magnetic field, we expect the harmonic spectrum of the coil to contain not only a uniform vertical mode, but also a determined set of non\hyphen   uniform modes that depend on the amount of symmetry-breaking caused by the geometrical imperfections included in the coil design. Finite-element simulations (COMSOL) of the field generated by the coil system, as well as later measurements with an automated field mapper,  yield a harmonic spectrum consistent with these expectations. 

The measured magnetic field satisfies the uniformity requirements of n2EDM. In particular, we showed that the problematic ``false EDM'' generated by the coil system was reproducible below the systematical requirement of $\qty{3e-28}{\ele.cm}$. Pushing the capabilities of the coil system even further, we used a dedicated set of coils to target problematic modes of the coil's harmonic spectrum in order to generate an even more uniform field. The resulting optimized field generates a false EDM strictly smaller than the systematical limit

Finally, the uniformity of the optimized magnetic field, which we characterize as the RMS deviation on its vertical component, is measured at $\sigma(B_z)=\qty{27}{pT}$ over the volume of the two precession chambers. The n2EDM coil system is thus able to generate a $\qty{1}{\micro T}$ vertical field with a relative root mean square deviation $\sigma(B_z)/B_z = \num{3e-5}$ inside a cylinder of radius $\qty{40}{cm}$ and height $\qty{30}{cm}$.

\vspace{10pt}

\textbf{Acknowledgments}

We acknowledge the mechanical department of the LPC Caen for the design (D.~Goupillière, Y.~Merrer) and construction (F.~Bourgeois, P.~Desrues, C.~Pain, B.~Bougard) of the precursor of the mapping robot which has been used in this work and of the coil system itself. K. \L{}ojek is also acknowledged for his indispensable work on the coils power supplies.

This project has received funding from
\begin{itemize}
\item The European Union under the ERC grant 716651-NEDM, the ERC grant 715031-Beam-EDM, and the Marie Skłodo\-wska-Curie grant agreement No 884104.  
\item The Swiss National Science Foundation grants \\ 
200020-188700, 200020-163413, 200011-178951, \\
200021-204118, 200021-172626, 206021-213222, \\ 169596 (PSI); \\
200020-215185, 200021-181996 (Bern); 
200441 (ETH); \\ and FLARE grants \\ 20FL21-186179, 20FL20-201473, 20FL20-216603. 
\item The French Agence Nationale de la Recherche (ANR) under grant ANR-14-CE33-0007.
\item The German Research Foundation (DFG) by the funding of the PTB core facility center of ultra-low magnetic field KO 5321/3-1 and TR 408/11-1 ; and by the Cluster of Excellence ‘Precision Physics, Fundamental Interactions, and Structure of Matter’ (PRISMA+ EXC 2118/1) funded within the German Excellence Strategy (Project ID 39083149)
\item The Fund for Scientific Research Flanders (FWO), and Project GOA/2010/10 of the KU Leuven.
\item The Polish National Science Center under grants \\ 
2015/18/M/ST2/00056, 2018/30/M/ST2/ 00319; \\ and the Polish Minister of Education and Science under the agreement No. 2022/WK/07. 
\item The School of Mathematical and Physical Sciences at the University of Sussex, as well as the STFC under grant ST/S000798/1.
\item  The Institute of Physics Belgrade through a grant by the Ministry of Education, Science and Technological Development of the Republic of Serbia.
\end{itemize}

\clearpage

\appendix

\section{The harmonic magnetic field expansion} \label{sec_harmonic}

\begin{table*}[ht!]
\centering
\begin{tabular}{@{} c c c c c @{}}
    \toprule
    $l$ &  $m$ & $\Pi_x$ & $\Pi_y$ & $\Pi_z$ \\
    \midrule
    $0$ & $-1$ & $0$ & $1$ & $0$ \\
    $0$ & $0$ & $0$ & $0$ & $1$ \\
    $0$ & $1$ & $1$ & $0$ & $0$ \\
    \midrule
    $1$ & $-2$ & $y$ & $x$ & $0$ \\
    $1$ & $-1$ & $0$ & $z$ & $y$ \\
    $1$ & $0$ & $-\frac{1}{2}x$ & $-\frac{1}{2}y$ & $z$ \\
    $1$ & $1$ & $z$ & $0$ & $x$ \\
    $1$ & $2$ & $x$ & $-y$ & $0$ \\
    \midrule
    $2$ & $-3$ & $2xy$ & $x^2-y^2$ & $0$ \\
    $2$ & $-2$ & $2yz$ & $2xz$ & $2xy$ \\
    $2$ & $-1$ & $-\frac{1}{2}xy$ & $-\frac{1}{4}\left(x^2+3y^2-4z^2\right)$ & $2yz$ \\
    $2$ & $0$ & $-xz$ & $-yz$ & $z^2-\frac{1}{2}(x^2+y^2)$ \\
    $2$ & $1$ & $-\frac{1}{4}\left(3x^2+y^2-4z^2\right)$ & $-\frac{1}{2}xy$ & $2xz$ \\
    $2$ & $2$ & $2xz$ & $-2yz$ & $x^2-y^2$ \\
    $2$ & $3$ & $x^2-y^2$ & $-2xy$ & $0$ \\
    \midrule
    3 & -4 & $3x^2y - y^3$ & $x^3 - 3xy^2$ & 0 \\
    3 & -3 & $6xyz$ & $3(x^2z - y^2z)$ & $3x^2y - y^3$ \\
    3 & -2 & $-\frac{1}{2}(3x^2y + y^3 - 6yz^2) $ & $-\frac{1}{2}(x^3 + 3xy^2 - 6xz^2)$ & $6xyz$ \\
    3 & -1 & $-\frac{3}{2}xyz$ & $-\frac{1}{4}(3x^2z + 9y^2z - 4z^3)$ & $3yz^2 -\frac{3}{4}(x^2y + y^3)$ \\
    3 &0 & $\frac{3}{8}(x^3 + xy^2 - 4xz^2)$ & $\frac{3}{8}(x^2y + y^3 - 4yz^2)$ & $z^3 - \frac{3}{2}z(x^2 + y^2)$ \\
    3 & 1 & $-\frac{1}{4}(9x^2z + 3y^2z - 4z^3)$ & $- \frac{3}{2}xyz$ & $3xz^2 - \frac{3}{4}(x^3 + xy^2)$ \\
    3 & 2 & $-x^3 + 3xz^2 $ & $- 3yz^2 + y^3$ & $3(x^2z - y^2z)$ \\
    3 & 3 & $3(x^2z - y^2z) $ & $- 6xyz$ & $x^3 - 3xy^2$ \\
    3 & 4 & $x^3 - 3xy^2 $ & $ - 3x^2y + y^3$ & $0$ \\
    \bottomrule
\end{tabular}
\caption{The basis of harmonic polynomials in Cartesian coordinates, sorted by degree up to $l=3$.}
\label{table_harmonics-cart}
\end{table*}

\begin{table*}[ht!]
\centering
\begin{tabular}{@{} c c c c c @{}}
    \toprule
    $l$ &  $m$ & $\Pi_\rho$ & $\Pi_\varphi$ & $\Pi_z$ \\
    \midrule
    0 & -1 & $\sin \varphi$ & $\cos \varphi$ & 0 \\
     0 & 0 & 0 & 0 & 1 \\
     0 & 1 & $\cos \varphi$ & $-\sin \varphi$ & 0 \\
     1 & -2 & $\rho \sin 2 \varphi$ & $\rho \cos 2 \varphi$ & 0 \\
     1 & -1 & $z \sin \varphi$ & $z \cos \varphi$ & $\rho \sin \varphi$ \\
     1 & 0 & $-\frac{1}{2} \rho$ & 0 & $z$ \\
     1 & 1 & $z \cos \varphi$ & $-z \sin \varphi$ & $\rho \cos \varphi$ \\
     1 & 2 & $\rho \cos 2 \varphi$ & $-\rho \sin 2 \varphi$ & 0 \\
     \midrule
     2 & -3 & $\rho^2 \sin 3 \varphi$ & $\rho^2 \cos 3 \varphi$ & 0 \\
     2 & -2 & $2 \rho z \sin 2 \varphi$ & $2 \rho z \cos 2 \varphi$ & $\rho^2 \sin 2 \varphi$ \\
     2 & -1 & $\frac{1}{4}\left(4 z^2-3 \rho^2\right) \sin \varphi$ & $\frac{1}{4}\left(4 z^2-\rho^2\right) \cos \varphi$ & $2 \rho z \sin \varphi$ \\
     2 & 0 & $-\rho z$ & 0 & $-\frac{1}{2} \rho^2+z^2$ \\
     2 & 1 & $\frac{1}{4}\left(4 z^2-3 \rho^2\right) \cos \varphi$ & $\frac{1}{4}\left(\rho^2-4 z^2\right) \sin \varphi$ & $2 \rho z \cos \varphi$ \\
     2 & 2 & $2 \rho z \cos 2 \varphi$ & $-2 \rho z \sin 2 \varphi$ & $\rho^2 \cos 2 \varphi$ \\
     2 & 3 & $\rho^2 \cos 3 \varphi$ & $-\rho^2 \sin 3 \varphi$ & 0 \\
     \midrule
     3 & -4 & $\rho^3 \sin 4 \varphi$ & $\rho^3 \cos 4 \varphi$ & 0 \\
     3 & -3 & $3 \rho^2 z \sin 3 \varphi$ & $3 \rho^2 z \cos 3 \varphi$ & $\rho^3 \sin 3 \varphi$ \\
     3 & -2 & $\rho\left(3 z^2-\rho^2\right) \sin 2 \varphi$ & $\frac{1}{2} \rho\left(6 z^2-\rho^2\right) \cos 2 \varphi$ & $3 \rho^2 z \sin 2 \varphi$ \\
     3 & -1 & $\frac{1}{4} z\left(4 z^2-9 \rho^2\right) \sin \varphi$ & $\frac{1}{4} z\left(4 z^2-3 \rho^2\right) \cos \varphi$ & $\rho\left(3 z^2-\frac{3}{4} \rho^2\right) \sin \varphi$ \\
     3 & 0 & $\frac{3}{8} \rho\left(\rho^2-4 z^2\right)$ & 0 & $\frac{1}{2} z\left(2 z^2-3 \rho^2\right)$ \\
     3 & 1 & $\frac{1}{4} z\left(4 z^2-9 \rho^2\right) \cos \varphi$ & $\frac{1}{4} z\left(3 \rho^2-4 z^2\right) \sin \varphi$ & $\rho\left(3 z^2-\frac{3}{4} \rho^2\right) \cos \varphi$ \\
     3 & 2 & $\rho\left(3 z^2-\rho^2\right) \cos 2 \varphi$ & $\frac{1}{2} \rho\left(\rho^2-6 z^2\right) \sin 2 \varphi$ & $3 \rho^2 z \cos 2 \varphi$ \\
     3 & 3 & $3 \rho^2 z \cos 3 \varphi$ & $-3 \rho^2 z \sin 3 \varphi$ & $\rho^3 \cos 3 \varphi$ \\
     3 & 4 & $\rho^3 \cos 4 \varphi$ & $-\rho^3 \sin 4 \varphi$ & 0 \\
     \midrule
     4 & 0 & $\frac{1}{2}\left(3 \rho^3 z-4 \rho z^3\right)$ & 0 & $\frac{1}{8}\left(8 z^4-24 \rho^2 z^2+3 \rho^4\right)$ \\
     \midrule
     5 & 0 & $\frac{5}{16}\left(-8 \rho z^4+12 \rho^3 z^2-\rho^5\right)$ & 0 & $\frac{1}{8}\left(8 z^5-40 \rho^2 z^3+15 \rho^4 z\right)$ \\
     \midrule
     6 & 0 & $\frac{3}{8} \rho\left(-8 z^5+20 \rho^2 z^3-5 \rho^4 z\right)$ & 0 & $\frac{1}{16}\left(16 z^6-120 \rho^2 z^4+90 \rho^4 z^2-5 \rho^6\right)$ \\
     \midrule
     7 & 0 & $\frac{7}{122} \rho\left(-64 z^6+240 \rho^2 z^4-120 \rho^4 z^2+5 \rho^6\right)$ & 0 & $\frac{1}{16} z\left(16 z^6-168 \rho^2 z^4+210 \rho^4 z^2-35 \rho^6\right)$ \\
    \bottomrule
\end{tabular}
\caption{The basis of harmonic polynomials in cylindrical coordinates, sorted by degree up to $l=7$. Only the $m=0$ modes are given for $l>3$.}
\label{table_harmonics-cyl}
\end{table*}

\begin{figure*}[ht!]
\hspace*{0cm}\centerline{
    \includegraphics[width=0.33\textwidth]{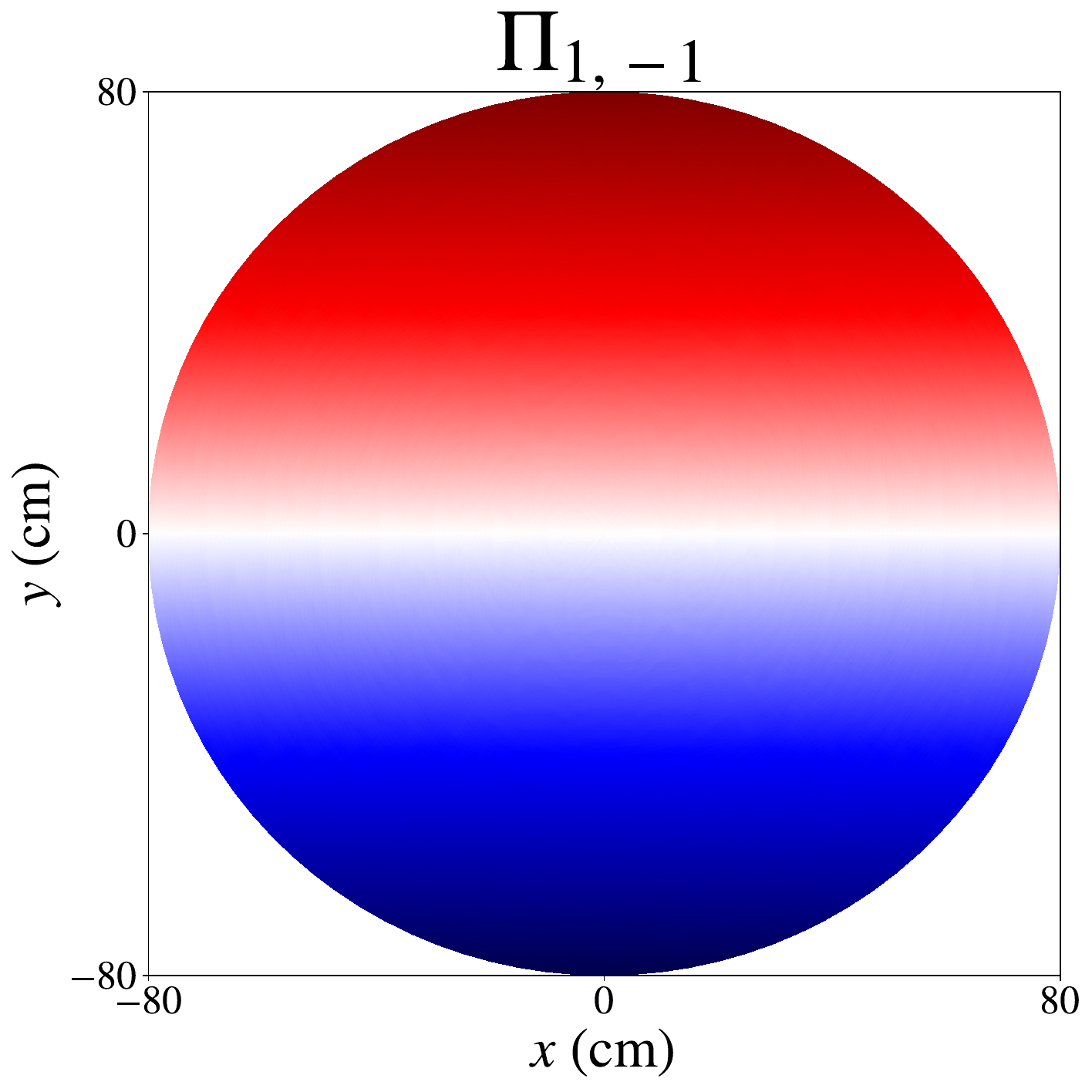}
    \includegraphics[width=0.33\textwidth]{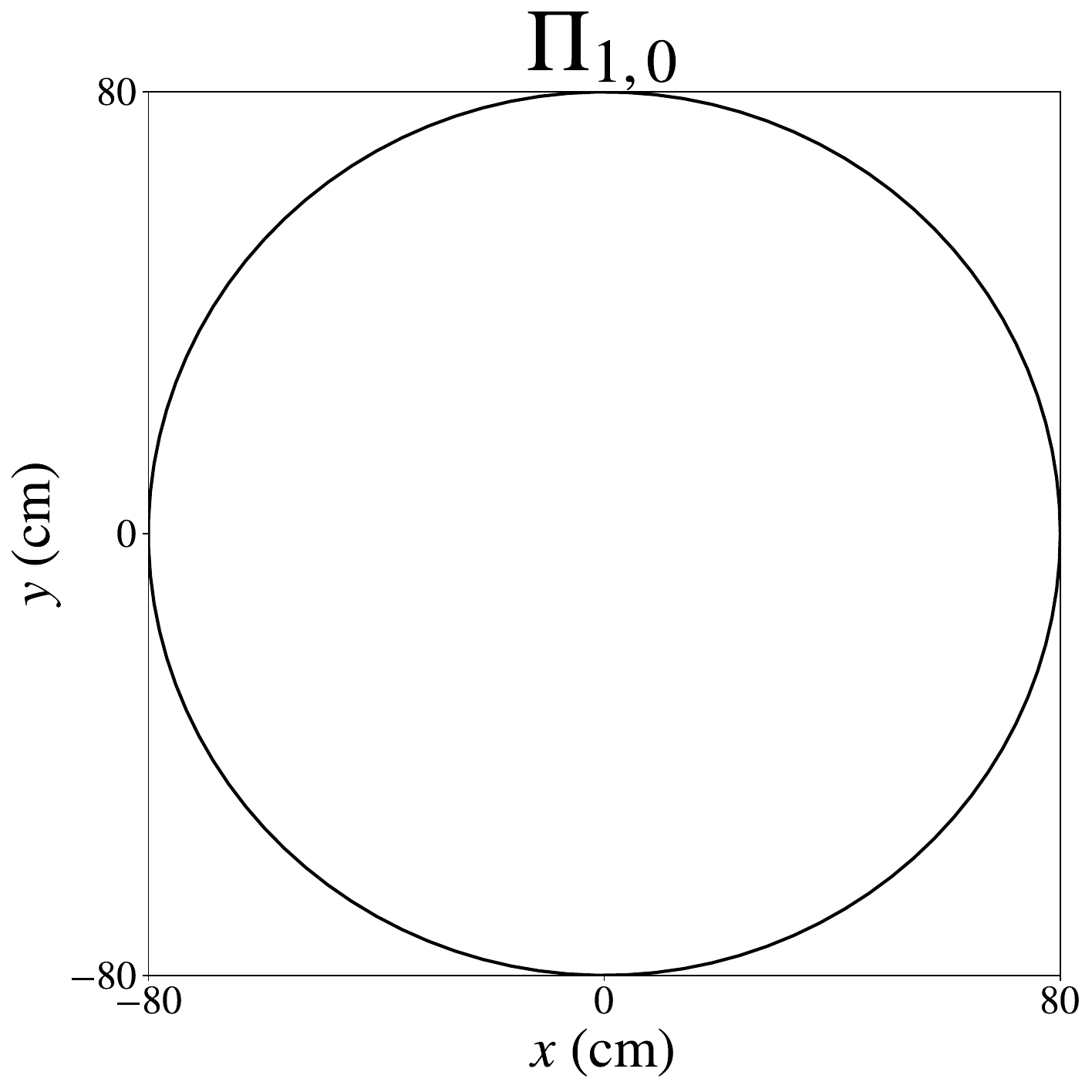}
    \includegraphics[width=0.33\textwidth]{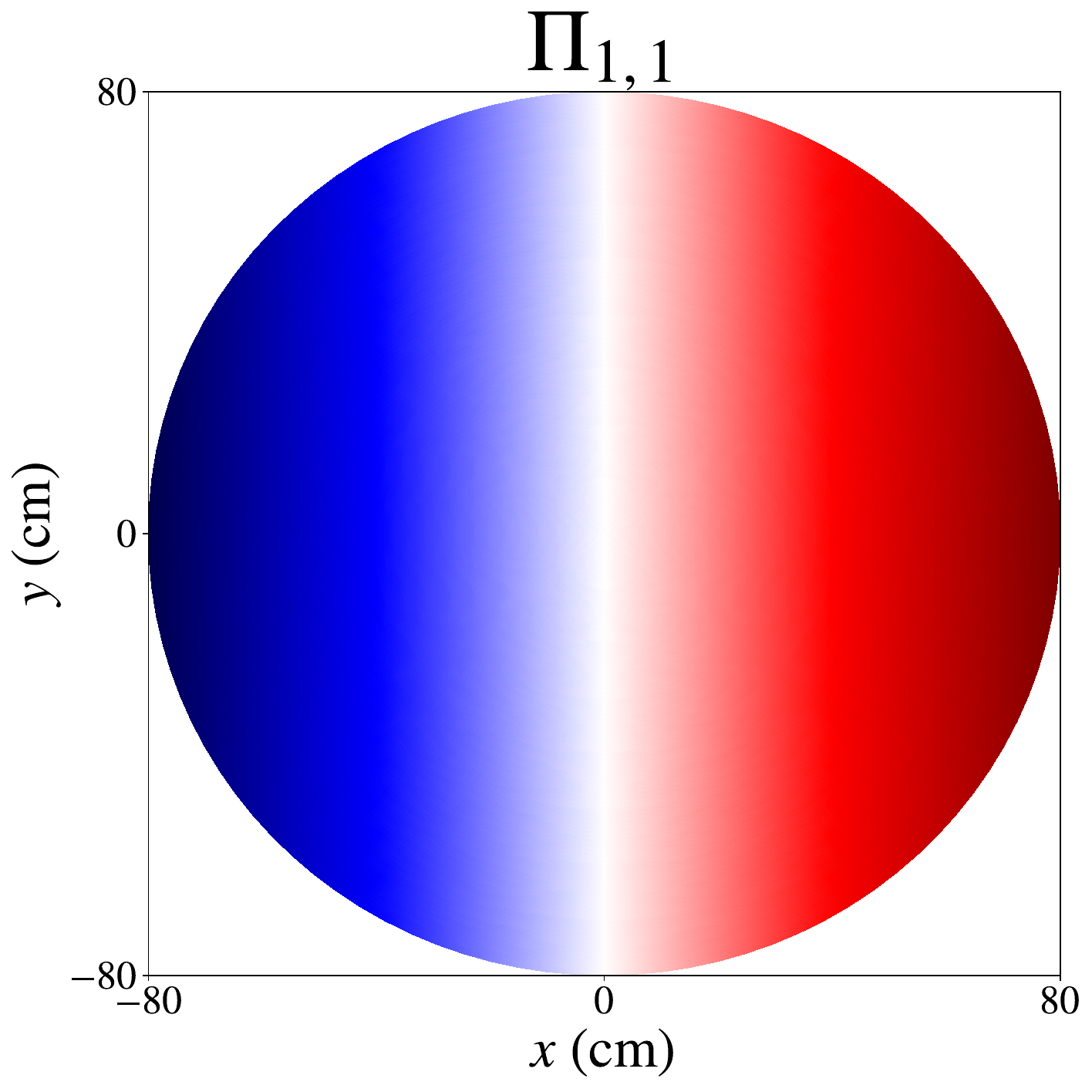}
}
\hspace*{0cm}\centerline{
    \includegraphics[width=0.33\textwidth]{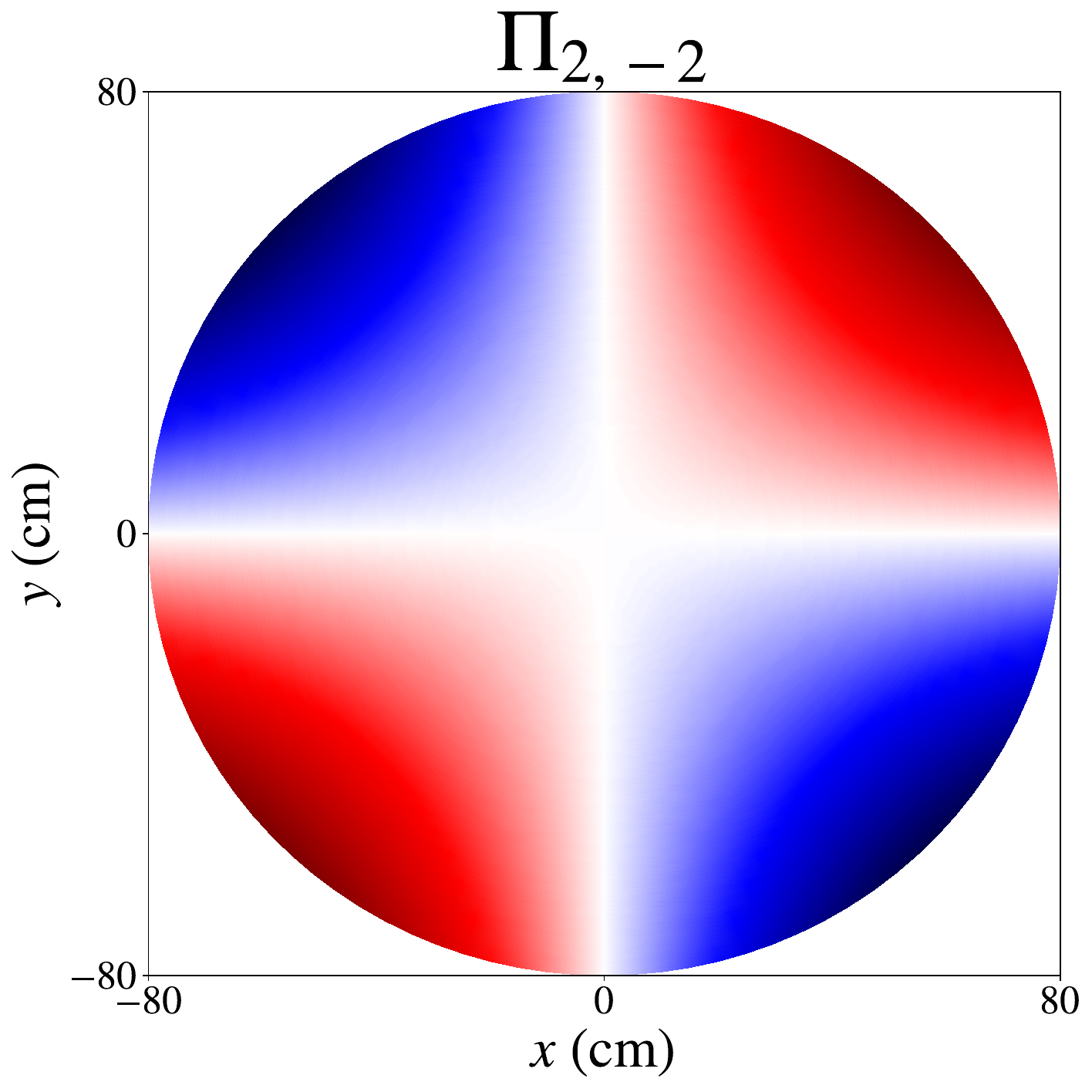}
    \includegraphics[width=0.33\textwidth]{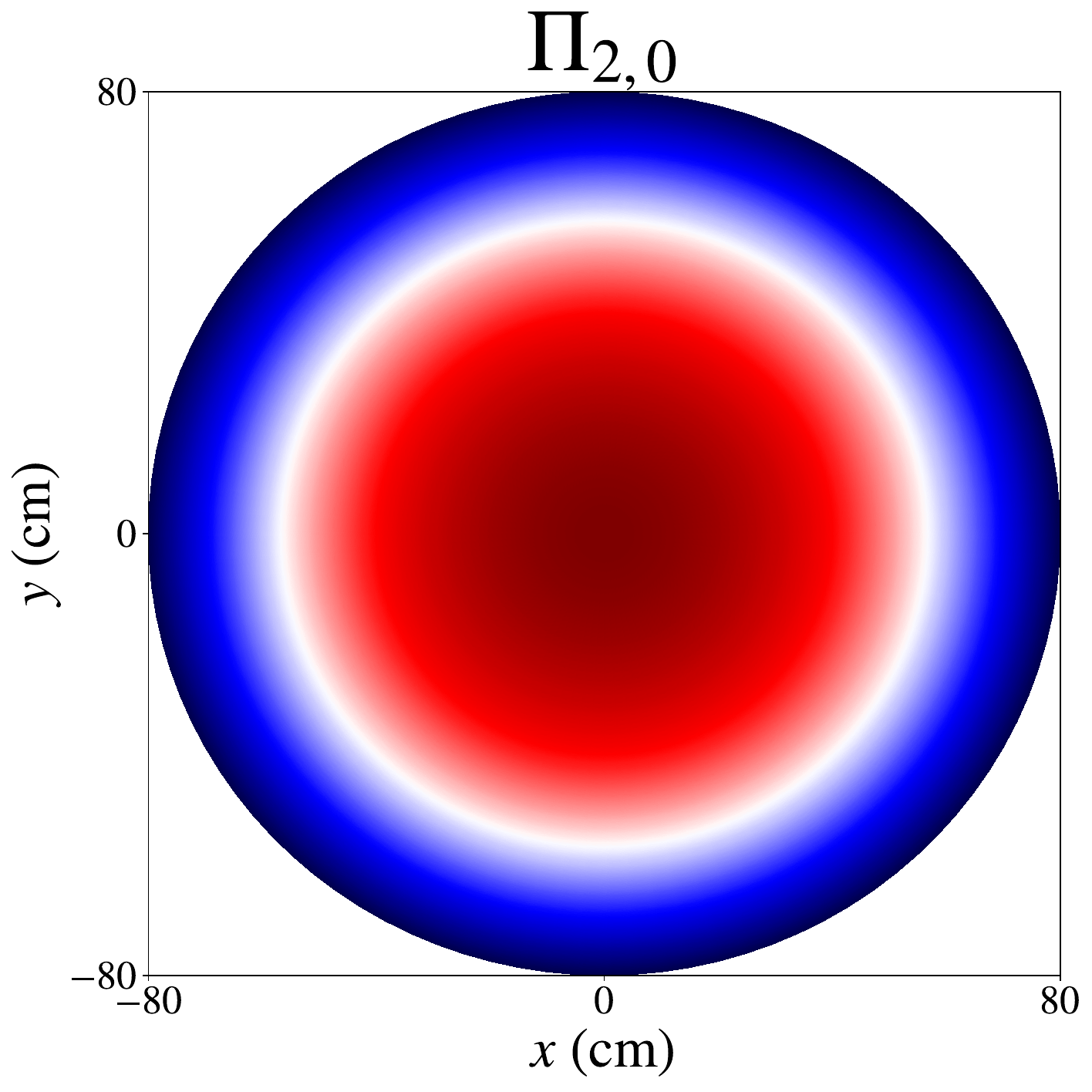}
    \includegraphics[width=0.33\textwidth]{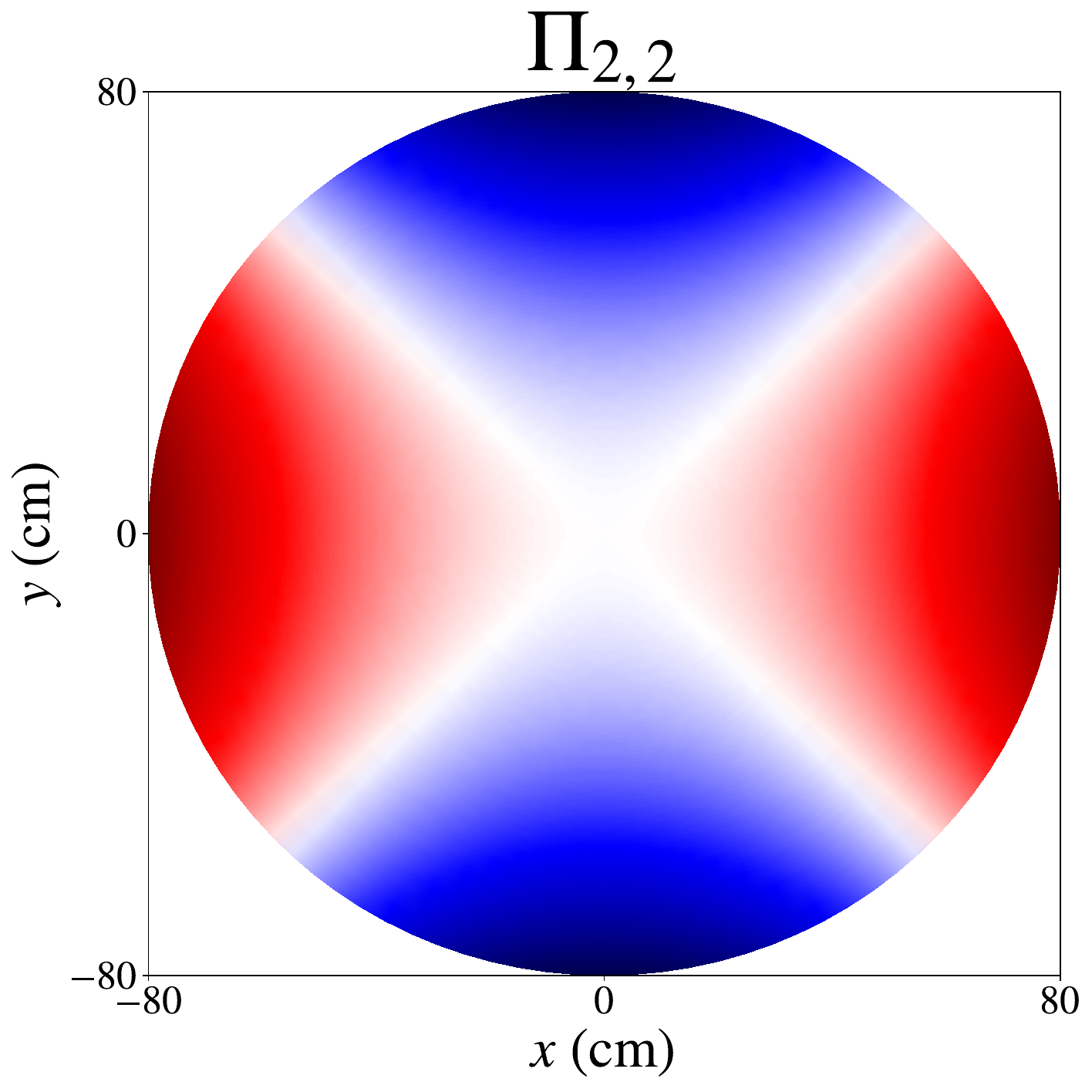}
    }
  \caption{Vertical field component of single harmonic modes in the $z=0$ plane. The field ranges in magnitude from blue to red.}
  \label{fig_harmonic-z0}
\end{figure*}

The harmonic expansion of the magnetic field given in equation \eqref{eq_harmonic-expansion} depends on a set of harmonic modes $\bm{\Pi}_{lm}(\mathbf{r})$ determined by solving Maxwell's equations in a region with no charge or magnetization
\begin{align}
    \boldsymbol{\nabla}\cdot \mathbf{B} &= 0, \label{eq_gradB} \\
    \boldsymbol{\nabla}\times \mathbf{B} &= 0. \label{eq_rotB}
\end{align}
Equation \eqref{eq_rotB} implies that the field can be written as the gradient of a potential $V$, with $\mathbf{B}=\boldsymbol{\nabla}V$. Equation \eqref{eq_gradB} imposes that this potential is a solution of Laplace's equation $\Delta V=0$, which in this case is expressed in the spherical coordinate system $(\rho, \theta, \varphi)$:
\begin{align}
     \frac{1}{\rho^2}\frac{\partial}{\partial \rho} \left(\rho^2 \frac{\partial V}{\partial \rho} \right) + \frac{1}{\rho^2\sin{\theta}}\frac{\partial}{\partial \theta} \left(\sin{\theta} \frac{\partial V}{\partial \theta} \right) \nonumber \\ + \frac{1}{\rho^2\sin^2{\theta}}\frac{\partial^2 V}{\partial \varphi^2} = 0. \label{eq_laplace-spherical}
\end{align}
The Laplace equation can be solved by separation of variables $ V(\rho,\theta,\varphi) = R(\rho)\Theta(\theta)\Phi(\varphi)$, which yields a set of solutions indexed by two integers $l$ and $m$, with $-l-1<m<l+1$. It was shown in \cite{uniformity} that, for some choice of normalization, these solutions can be written as
\begin{align}
    V_{lm}(\rho, \theta, \varphi) = &\frac{(l-1)!(-2)^{|m|}}{(l+|m|)!} \rho^lP_l^{|m|}(\cos(\theta)) \times \nonumber \\
    &\begin{cases}
        \cos{(|m|\varphi)} &\text{ if } m\geq 0 \\
        \sin{(|m|\varphi)} &\text{ if } m< 0
    \end{cases}, \label{eq_magneticpotential}
\end{align}
where the $P_l^m$ are the associated Legendre polynomials. The $l$-degree harmonic polynomials are finally determined by differentiation of the $l+1$-degree field potential:
\begin{equation}
    \boldsymbol{\Pi}_{lm}(\mathbf{r}) = \boldsymbol{\nabla} V_{l+1,m}(\mathbf{r}). \label{eq_gradphi}
\end{equation}
In Cartesian coordinates, the $\boldsymbol{\Pi}_{lm}$ are $l$-degree polynomial functions of $x,y,z$, given up to order $l=3$ in \ref{table_harmonics-cart}, where $\boldsymbol{\Pi}_{lm} = \Pi_{x,lm}\mathbf{e}_x + \Pi_{y,lm}\mathbf{e}_y + \Pi_{z,lm}\mathbf{e}_z$. 
To obtain the harmonic functions in cylindrical coordinates, we let 
\begin{align}
    \Pi_{\rho,lm} &= \cos{(m\varphi)}\Pi_{x,lm} + \sin{(m\varphi)}\Pi_{y,lm}, \\
    \Pi_{\varphi,lm} &= -\sin{(m\varphi)}\Pi_{x,lm} + \cos{(m\varphi)}\Pi_{y,lm},
\end{align}
so that $\boldsymbol{\Pi}_{lm} = \Pi_{\rho,lm}\mathbf{e}_\rho + \Pi_{\varphi,lm}\mathbf{e}_\varphi + \Pi_{z,lm}\mathbf{e}_z$. Furthermore, we usually separate the polynomial and angular functions and write
\begin{equation}
        \bm{\Pi}_{lm}(\mathbf{r}) = \bm{\Tilde{\Pi}}_{lm}(\rho,z) \cdot \mathbf{y}_{lm}(\varphi),
\end{equation}
where the reduced harmonic functions $\bm{\Tilde{\Pi}}_{lm}(\rho,z)$ are polynomials in $\rho,z$, and
\begin{equation}
   \mathbf{y}_{lm}(\varphi) =
    \begin{cases}
        \cos{(m\varphi)}\mathbf{e}_\rho + \sin{(m\varphi)}\mathbf{e}_\varphi + \cos{(m\varphi)}\mathbf{e}_z &\text{if } m\geq 0, \\
        \sin{(m\varphi)}\mathbf{e}_\rho + \cos{(m\varphi)}\mathbf{e}_\varphi + \sin{(m\varphi)}\mathbf{e}_z &\text{if } m< 0.
    \end{cases}
\end{equation}
These cylindrical harmonic functions are given up to order $l=7$ in table \ref{table_harmonics-cyl}.
A few useful harmonic modes are represented in the $z=0$ transverse plane in figure \ref{fig_harmonic-z0}.


\section{Phantom modes and magnetic gradient normalization} \label{sec_norm}

We recall here the explicit expressions of the two contributions to the false EDM in the geometry of n2EDM, and give a useful normalization of the generalized gradients.

The top-bottom magnetic gradient existing between the two chambers of n2EDM is defined in \cite{n2edm} as
\begin{align}
   G_{\text{TB}} &= \frac{\left<B_z\right>_\text{TOP} - \left<B_z\right>_\text{BOT} }{H'} \nonumber \\
   &= G_{1\,0} - L_3^2G_{3\,0} + L_5^4G_{5\,0} - \hdots \label{eq_GTBdef},
\end{align}
where $H'=\qty{18}{cm}$ is the height difference between the centers of the two chambers, and where the second line is obtained by using the harmonic expansion of $B_z$. The $L_l$ are then geometric coefficients with a unit of distance, provided in table \ref{tab_geometric-coefs}.

\begin{table*}[hb!]
\centering
    \begin{tabular}{@{\kern\tabcolsep} l c c @{\kern\tabcolsep}}
    \toprule
    Coef. & \textbf{Expression} & \textbf{Value} \\
    \midrule
        $L_3^2$ & $\frac{3R^2}{4} - \frac{H^2 + H'^2}{4}$ & $(\qty{32.9}{cm})^2$  \\
        $L_5^4$ & $\frac{5R^4}{8} - \frac{5R^2(H^2+H'^2)}{8} + \frac{3H^4+10H^2H'^2+3H'^4}{48} $ & $(\qty{32.7}{cm})^4$ \\
        $L_7^6$ & $\frac{1}{16}\left(\frac{35R^6}{4} - \frac{70R^4(H^2 + H'^2)}{3} + \frac{21R^2(3H^4 + 10H^2H'^2 + 3H'^4)}{12} - \frac{H^6 + 7 H^4H'^2 + 7 H^2H'^4 + H'^6}{4}\right)$ & $(\qty{32.5}{cm})^2$ \\
    \bottomrule
    \end{tabular}
    \caption{Geometrical coefficient $L_{2k+1}^{2k}$ for each $2k+1$ term of the top-bottom gradient expansion \eqref{eq_GTBdef}, up to order $2k+1=7$.}
    \label{tab_geometric-coefs}
\end{table*}

The false EDM in the n2EDM geometry expressed by equation \eqref{eq_falseEDM-gradients} is generated by $l$-odd, $m=0$ harmonic modes, some of which also generate a top-bottom gradient. The rest, which satisfy $G_\text{TB}=0$, are the so-called phantom modes. These can be written as
\begin{equation}
    \Acute{\boldsymbol{\Pi}}_{2k+1,0} = \frac{L_{2k+1}^{2k}}{D_{2k+1}^{2k}}\left[\boldsymbol{\Pi}_{1\,0} - \frac{(-1)^k}{L_{2k+1}^{2k}}\boldsymbol{\Pi}_{2k+1,0}\right]. \label{eq_phantoms-expression}
\end{equation}
The geometric coefficents $D_{l}$ are determined by the normalization condition $\left<\rho\Acute{\Pi}_{l0}^{(\rho)}\right>_\text{TOP} = -R^2/4$.
For odd-order modes, this yields
\begin{equation}
    D_{2k+1}^{2k} = \left[ L_{2k+1}^{2k} - (-1)^k \frac{\left<\rho\Pi_{2k+1,0}^{(\rho)}\right>}{-R^2/4} \right].
\end{equation}
For even degree modes, which do not generate a top-bottom gradient, we simply obtain
\begin{equation}
    D_{2k}^{2k-1} = \frac{\left<\rho \Pi_{2k,0}^{(\rho)}\right>_{\text{TOP}}}{-R^2/4}.
\end{equation}
Their numerical values for $l\leq7$ are given in table \ref{table_D}. Combining the above, a magnetic field of the form
\begin{equation}
    \mathbf{B} = G_{\text{TB}}\boldsymbol{\Pi}_{1\,0} + \Acute{G}_{3\, 0}\Acute{\boldsymbol{\Pi}}_{3\, 0} + \Acute{G}_{5\, 0} \Acute{\boldsymbol{\Pi}}_{5\, 0} + \hdots, \label{eq_Bphantom}
\end{equation}
with the normalized gradients $\Acute{G}_{l} = D^{l-1}_{l} G_{l}$, generates precisely the false EDM given by equation \eqref{eq_falseEDM-gradients}.

\section{Symmetries of the coil system and allowed harmonic modes} \label{sec_symmetry}

Here we determine the symmetries of several geometrical configurations of the n2EDM coil system, and derive the harmonic shape of the magnetic field that preserves these symmetries.

The current system described by the $B_0$ coil current loops, pictured in \ref{fig_B0_coil_design}, and the innermost layer of the MSR, is invariant to a factor $\pm 1$ under a number of spatial transformations that depends on the inclusion of some symmetry-breaking features in the coil's design. Because the magnetic field induced by a current running through a solenoid is a pseudo-vector while the current field is a vector, we know that the symmetries (factor $1$ transformations) of the coil's current system are anti-symmetries of the induced magnetic field, and that its anti-symmetries (factor $-1$ transformations) correspond to magnetic field symmetries. After identifying the symmetries and anti-symmetries of the coil system, we determine which modes of the harmonic expansion are anti-symmetric and symmetric in the coil's geometry, and are therefore allowed by the coil's design. These results are featured in table \ref{tab_allowed-modes}.

We explain this process for an idealized $B_0$ coil, which consists of a perfect square solenoid where none of the symmetry-breaking features represented in red, green, and blue in figure \ref{fig_B0_coil_design} are considered. The current system of the idealized coil is invariant to a factor $\pm 1$ under a set of $16$ spatial transformations
\begin{equation}
    \mathcal{D}_{4h} = \{I, P, \sigma_x, \sigma_y, \sigma_z, R_x^2, R_y^2, R_z^2, R_z, R_z^{-1}, \sigma_{xy}, \sigma_{-xy}, R_z', R_z^{\prime-1}, \sigma_{xy}', \sigma_{-xy}' \}.
\end{equation}
These are invertible $3\times 3$ matrices, which together with matrix multiplication possess the mathematical structure of a group. This group can be generated by three of its elements, one of the two vertical plane reflections $\sigma_x, \sigma_y$, the horizontal plane reflection $\sigma_z$, and the $\pi/2$-rotation around the vertical axis $R_z$, such that $\mathcal{D}_{4h} = \left<\sigma_x, \sigma_z, R_z \right>$. We give the explicit forms of these generators:
\begin{align}
    \sigma_x &= 
    \begin{pmatrix}
    -1 & 0 & 0 \\
    0 & 1 & 0 \\
    0 & 0 & 1 \\
    \end{pmatrix} 
    & 
    \sigma_y &= 
    \begin{pmatrix}
    1 & 0 & 0 \\
    0 & -1 & 0 \\
    0 & 0 & 1 \\
    \end{pmatrix}
    &
    \sigma_z &= 
    \begin{pmatrix}
    1 & 0 & 0 \\
    0 & 1 & 0 \\
    0 & 0 & -1 \\
    \end{pmatrix}
    &
    \\
    & &
    R_z &= 
    \begin{pmatrix}
    0 & -1 & 0 \\
    1 & 0 & 0 \\
    0 & 0 & 1 \\
    \end{pmatrix}, \label{eq_generators}
\end{align}
while the remaining group elements can be found through the following combinations:
\begin{align}
    I &= \sigma_x^2=\sigma_y^2=\sigma_z^2 & P &= \sigma_x\sigma_y\sigma_z \nonumber \\
    R_x^2 &= \sigma_y\sigma_z & R_y^2 &= \sigma_x\sigma_z & R_y^2 &= \sigma_x\sigma_y \nonumber \\
    \sigma_{xy} &= \sigma_x R_z & \sigma_{-xy} &= \sigma_y R_z \nonumber \\
    R_z' &= \sigma_z R_z & \sigma_{xy}' &= \sigma_z \sigma_{xy} & \sigma_{-xy}' &= \sigma_z \sigma_{-xy} 
\end{align}

In more specific terms, the current system is transformed by a representation, hereafter referred to as the \textit{current representation}, of the $\mathcal{D}_{4h}$ group
\begin{equation}
    \rho_c : \mathcal{D}_{4h} \longrightarrow GL(V_c),
\end{equation}
where $GL$ is the general linear group, $V_c$ is the vector space where the current field lives, and where the elements of $\rho_c(\mathcal{D}_{4h})$ satisfy by definition
\begin{equation}
    \rho_c(T_1T_2)=\rho_c(T_1)\rho_c(T_2),\quad \forall T_1, T_2 \in \mathcal{D}_{4h}. \label{eq_representation}
\end{equation}
The elements $\rho_c(T)$ are determined by considering that these are linear transformations of the current field $\mathbf{I}(\mathbf{r})$, given by
\begin{align}
    \rho_c(T) : \mathbf{I}(\mathbf{r}) \longmapsto \rho_c(T)\mathbf{I}(\mathbf{r}) &= T\mathbf{I}(T^{-1}\mathbf{r}). \label{eq_coil-representation}
\end{align}
Solving the above for $T=\sigma_x, \sigma_z, R_z$, which are generators of $\mathcal{D}_{4h}$, yields the idealized coil character given in table \ref{tab_allowed-modes} and completely determines the symmetry of the idealized coil system. For instance, applying the reflection symmetry w.r.t. the YZ plane $\sigma_x$ to the idealized solenoid $B_0$ coil in equation \eqref{eq_coil-representation}, using $\sigma_x$ from equation \eqref{eq_generators}, leads to $\rho_c{(\sigma_x)}=-1$.

The shape of the magnetic field that preserves this symmetry is obtained by considering the \textit{magnetic representation} $\rho_b$, whose elements are linear transformations of the magnetic field such that
\begin{align}
    \rho_b(T) : \mathbf{B}(\mathbf{r}) \longmapsto \rho_b(T)\mathbf{B}(\mathbf{r}) &= \det(T)T\mathbf{B}(T^{-1}\mathbf{r}). 
\end{align}
Requiring that the symmetry of the magnetic representation matches the symmetry of the current representation, with $\rho_b(\sigma_x)=-1$, $\rho_b(\sigma_z)=1$, and $\rho_b(R_z)=1$, amounts to looking for a field that satisfies the three following equations simultaneously:
\begin{align}
    \begin{pmatrix}
        B_x \\
        B_y \\
        B_z 
    \end{pmatrix}(x, y, z)
    &=
    \begin{pmatrix}
        -B_x \\
        B_y \\
        B_z 
    \end{pmatrix}(-x, y, z) \label{eq_sx}
    \\
    \begin{pmatrix}
        B_x \\
        B_y \\
        B_z 
    \end{pmatrix}(x, y, z)
    &=
    \begin{pmatrix}
        -B_x \\
        -B_y \\
        B_z 
    \end{pmatrix}(x, y, -z) \label{eq_sz}
    \\
    \begin{pmatrix}
        B_x \\
        B_y \\
        B_z 
    \end{pmatrix}(x, y, z)
    &=
    \begin{pmatrix}
        -B_y \\
        B_x \\
        B_z 
    \end{pmatrix}(y, -x, z) \label{eq_rz}
\end{align}
Making use of the harmonic field expansion \eqref{eq_harmonic-expansion}, we identify the harmonic modes that satisfy these equations and thus preserve the idealized coil symmetry. These are referred to as \textit{allowed} modes and are given by table \ref{tab_allowed-modes}. We then see that harmonic modes $\boldsymbol{\Pi}_{2\,0}, \boldsymbol{\Pi}_{40}, \boldsymbol{\Pi}_{44}, \hdots$ are allowed by the idealized coil geometry.

The harmonic modes allowed by less ideal thus more restrictive coil geometries are determined in the same fashion and featured in the same table, by considering subgroups of $\mathcal{D}_{4h}$. Incorporating the vacuum pipes and neutron guides in the coil geometry (in green in figure \ref{fig_B0_coil_design}) breaks the idealized coil symmetry $\mathcal{D}_{4h}$ as it is no more invariant under the $\pi/2$-rotation around the vertical axis. The symmetry of this hole-broken coil is given by the $\mathcal{D}_{2h}$ group, the largest subgroup of $\mathcal{D}_{4h}$ which does not contain $R_z$. Similarly, the $\mathcal{C}_{2v(x)}$, $\mathcal{C}_{2v(y)}$, and $\mathcal{C}_{2v(z)}$ subgroups are obtained by removing $\sigma_x$, $\sigma_y$, and $\sigma_z$ respectively from $\mathcal{D}_{2h}$. $\mathcal{C}_{2v(y)}$ in particular corresponds to a coil symmetry broken by the MSR door (in blue in figure \ref{fig_B0_coil_design}).

\begin{table*}[ht!]
    \ra{1.3}
    \centering
     \begin{tabular}{@{\kern\tabcolsep} l c c c c c c @{\kern\tabcolsep}}
     \toprule
        \textbf{Coil geometry} & \textbf{Irrep(Group)} & \multicolumn{4}{c}{\textbf{Coil character}} & \textbf{Allowed modes} \\
        \cmidrule(lr){3-6}
        & & $\sigma_x$ & $\sigma_y$ & $\sigma_z$ & $R_z$ & \\
        \midrule
        Solenoid & $\rho_c(\mathcal{D}_{4h})$ & $-1$ & $-1$ & $1$ & $1$ & $\{\boldsymbol{\Pi}_{2k,4n}\}_{k,n\in\mathbb{N}}$ \\
        w/ tubes or guides & $\rho_c(\mathcal{D}_{2h})$ & $-1$ & $-1$ & $1$ & broken & $\{\boldsymbol{\Pi}_{2k,2n}\}_{k,n\in\mathbb{N}}$ \\
        w/ $x$-shift & $\rho_c(\mathcal{C}_{2v(x)})$ & broken & $-1$ & $1$ & broken & $\{\boldsymbol{\Pi}_{2k,2n},\boldsymbol{\Pi}_{2k+1, 2n+1}\}_{k,n\in\mathbb{N}}$\\
        w/ $y$-shift or door & $\rho_c(\mathcal{C}_{2v(y)})$ & $-1$ & broken & $1$ & broken & $\{\boldsymbol{\Pi}_{2k,2n},\boldsymbol{\Pi}_{2k+1, -2n-1}\}_{k,n\in\mathbb{N}}$ \\
        w/ $z$-shift & $\rho_c(\mathcal{C}_{2v(z)})$ & $-1$ & $-1$ & broken & broken & $\{\boldsymbol{\Pi}_{2k,2n},\boldsymbol{\Pi}_{2k+1, 2n}\}_{k,n\in\mathbb{N}}$ \\
    \bottomrule
    \end{tabular}
    \caption{Augmented character table of several symmetry groups of the coil system. The leftmost column specifies a more or less ideal coil geometry. The second column gives the coil representation of a symmetry group $\mathcal{G}$ satisfied by this geometry. The following four columns provide the representation $\rho_c(T)$ of a given symmetry $T\in\mathcal{G}$, which altogether constitute the character of $\rho_c(\mathcal{G})$. The rightmost columns specifies the harmonic modes whose magnetic representation $\rho_b(\mathcal{G})$ shares the same character as the coil representation, and are therefore allowed by the corresponding coil geometry.}
    \label{tab_allowed-modes}
\end{table*}

\section{Geometrical description of the gradient coils}
\label{Gradient_coils}

The following figures present the design of the seven gradient coils built for the experiment. The $G_{10}$, $G_{20}$ and $G_{30}$ coils are fixed on the $B_0$ coil support while all others are fixed on an additional support attached to the $B_0$ one. The scale unit showed on the cube sides is the meter. The axis origin is at the center of the system. The $B_0$ door is schemed in the front face by the black parallelogram. The blue pattern corresponds the current path, the red arrows give the direction of the current flow. All coils have their own power supply.

\onecolumn
\begin{figure}[ht!]
\begin{center} 
\includegraphics[width=0.76\textwidth]{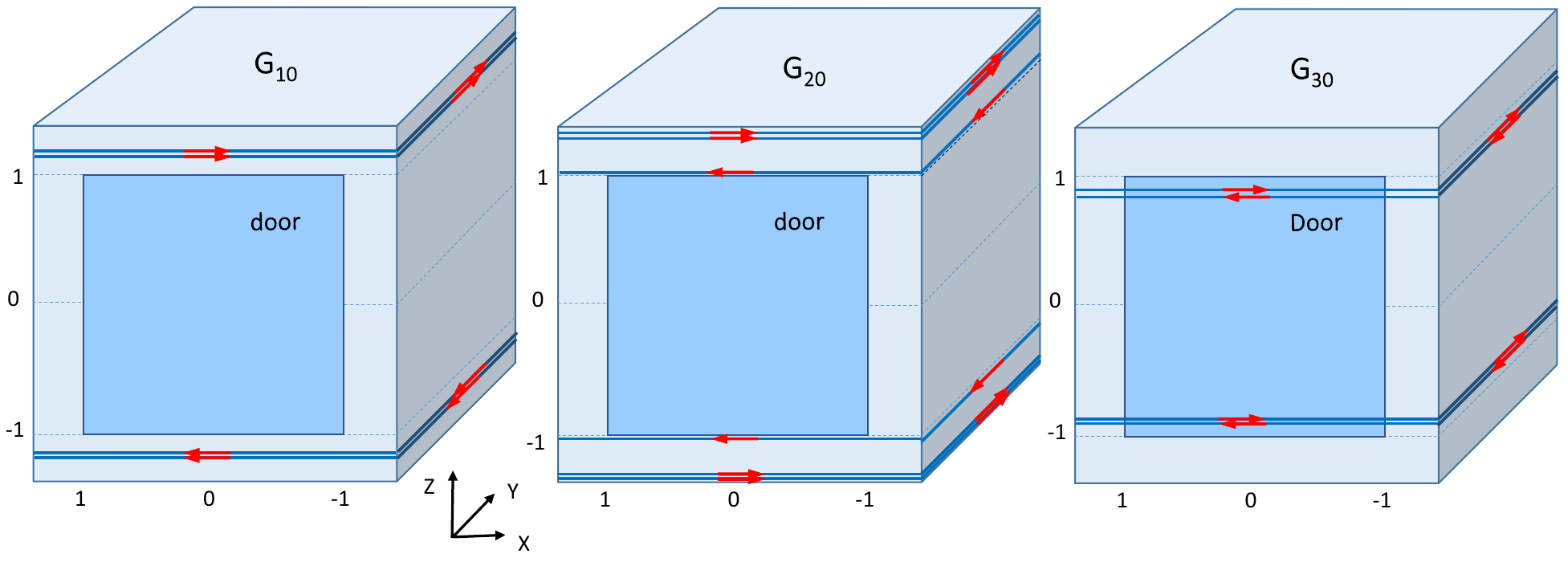} 
\caption{Left: $G_{10}$ gradient coil. The coil is made of four horizontal  square loops located at z = $\pm$ 1335~mm and z = $\pm$ 1035~mm. Middle: $G_{20}$ gradient coil. The $G_{20}$ coil is made of six horizontal square loops located at z = $\pm$ 1335~mm, z = $\pm$ 1035~mm and z = $\pm$ 1005~mm. Right: $G_{30}$ gradient coil. The $G_{30}$ coil is made of four horizontal square loops located at z = $\pm$ 937.5~mm, z = $\pm$ 1035~mm and z = $\pm$ 892.5~mm.  }
\label{}
\end{center}
\end{figure} 

\vspace{-1 cm}
\begin{figure}[ht!]
\begin{center}
  \centering 
  \includegraphics[trim = 2.5cm 3.5cm 3.5cm 3.5cm, clip, scale=0.4]{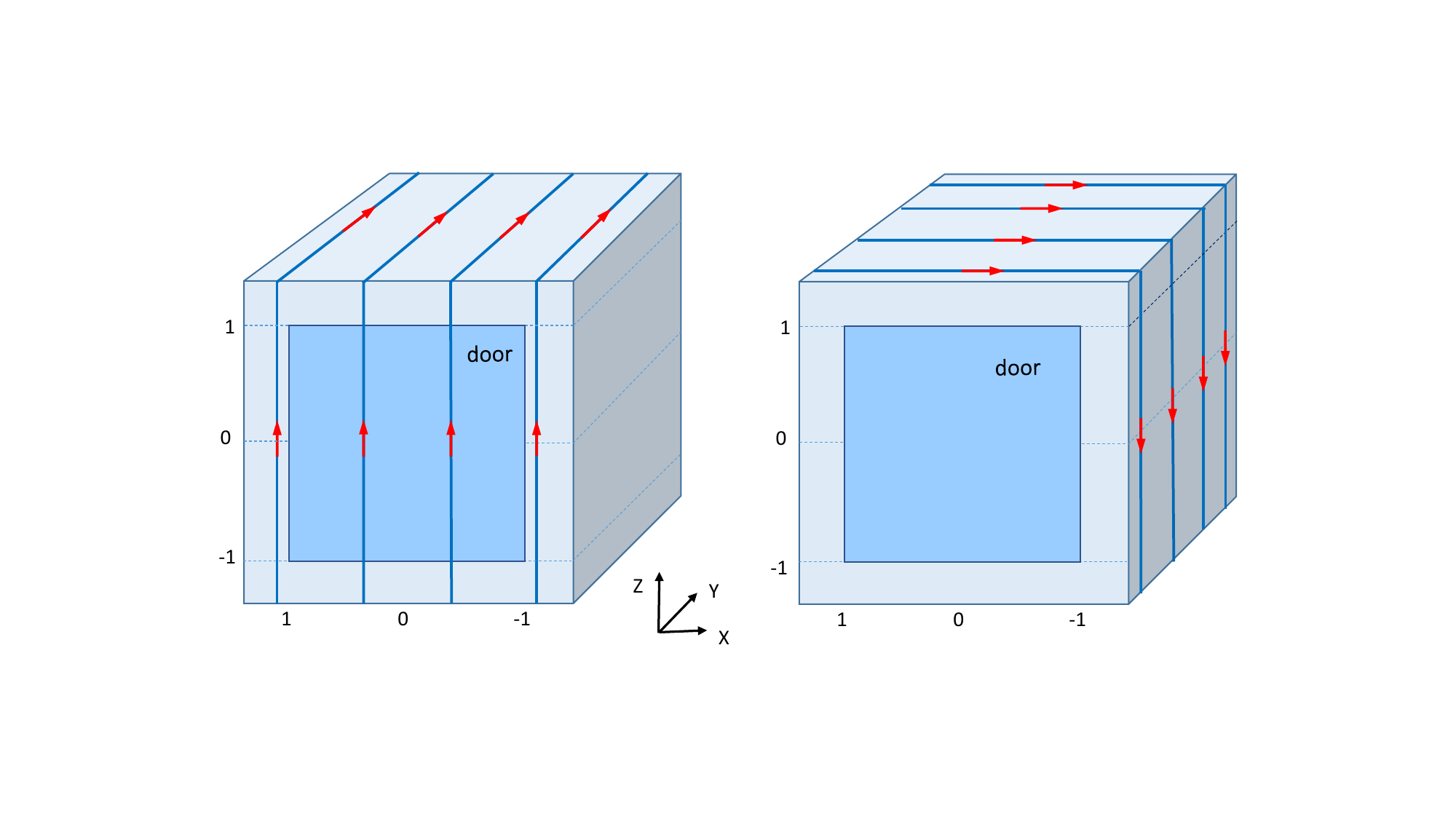}
  \caption{Design of the $G_{01}$ (left) and $G_{0-1}$ (right) gradient coils. The $G_{01}$ coil is made of four vertical square loops located at x = $\pm$ 1118~mm and x = $\pm$ 359~mm. The $G_{01}$ coil is made of four vertical square loops located at y = $\pm$ 1118~mm and y = $\pm$ 359~mm.}
\label{}
\end{center}
\end{figure}

\vspace{-1.5 cm}
\begin{figure}[ht!]
\begin{center}
  \includegraphics[trim = 5.5cm 4.cm 16.5cm 4.cm, clip, scale=0.4]{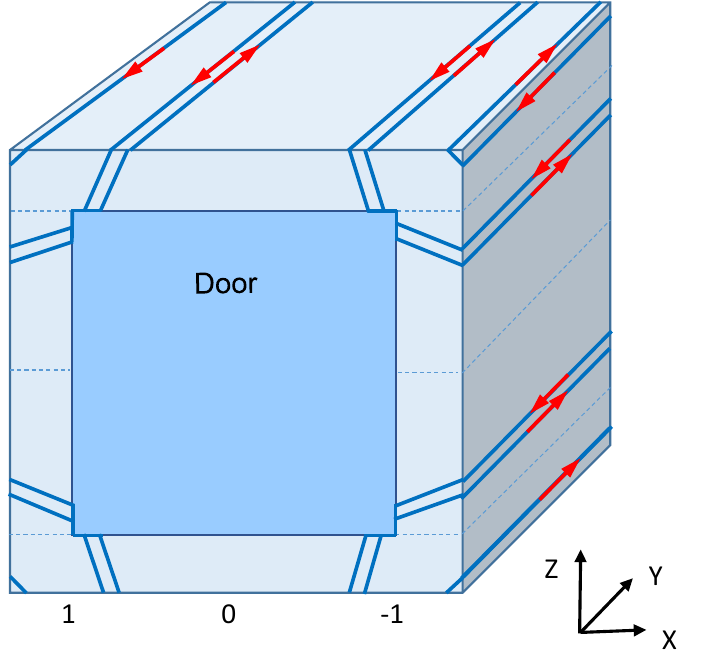}
\caption{Design of the $G_{11}$ gradient coil. The $G_{1-1}$ gradient coil has exactly the same design rotated by 90 degrees i.e. symmetric with respect to (Ox,Oz) instead of (Oy,Oz).}
\label{}
\end{center}
\end{figure}

\twocolumn

\printbibliography

\end{document}